\documentclass[12pt,aps,prd,preprint,tightenlines,showpacs,nofootinbib,
superscriptaddress]{revtex4-1}
\newcommand{\PRE}[1]{{#1}} 

\usepackage{hyperref}      
\usepackage{bm} 
\usepackage{amsmath} 
\usepackage{epsfig}
\usepackage{subfigure}
\usepackage{dsfont}

\newcommand{\mweak}{m_{\text{weak}}}

\newcommand{\ifb}{\text{fb}^{-1}}

\newcommand{\mev}{\text{MeV}}
\newcommand{\gev}{\text{GeV}}
\newcommand{\tev}{\text{TeV}}

\newcommand{\cm}{\text{cm}}

\renewcommand{\eqref}[1]{Eq.~(\ref{#1})}
\newcommand{\eqsref}[2]{Eqs.~(\ref{#1}) and (\ref{#2})}

\newcommand{\secref}[1]{Sec.~\ref{sec:#1}}

\newcommand{\figref}[1]{Fig.~\ref{fig:#1}}
\newcommand{\figsref}[2]{Figs.~\ref{fig:#1} and \ref{fig:#2}}
\newcommand{\Figref}[1]{Figure~\ref{fig:#1}}

\newcommand{\mchi}{m_{\chi}}

\newcommand{\Bino}{\tilde{B}}
\newcommand{\Wino}{\tilde{W}}
\newcommand{\Higgsino}{\tilde{H}}

\newcommand{\mgaugino}{M_{1/2}}

\newcommand{\Omegachi}{\Omega_{\Neut}}

\newcommand{\Neut}{\chi}

\newcommand{\sigmaSI}{\sigma^{\text{SI}}}

\newcommand{\ah}{a_{\Higgsino}}
\newcommand{\ahu}{a_{\Higgsino_u}}
\newcommand{\ahd}{a_{\Higgsino_d}}

\newcommand{\ab}{a_{\Bino}}
\newcommand{\aw}{a_{\Wino}}
\newcommand{\phiCP}{\phi_{\text{CP}}}
\newcommand{\disabled}[1]{{}}

\begin{document}

\preprint{UCI-TR-2011-28}

\title{ \PRE{\vspace*{0.5in}}
Focus Point Supersymmetry Redux
\PRE{\vspace*{0.3in}} }

\author{Jonathan L.~Feng}
\affiliation{Department of Physics and Astronomy, University of
California, Irvine, California 92697, USA
\PRE{\vspace*{.2in}}
}

\author{Konstantin T.~Matchev}
\affiliation{Department of Physics, University of Florida, Gainesville,
  Florida 32611, USA
\PRE{\vspace*{.5in}}
}

\author{David Sanford\PRE{\vspace*{.4in}}}
\affiliation{Department of Physics and Astronomy, University of
California, Irvine, California 92697, USA
\PRE{\vspace*{.2in}}
}

\date{March 20, 2012}
\PRE{\vspace*{0.6in}}

\begin{abstract}
\PRE{\vspace*{.3in}} Recent results from Higgs boson and supersymmetry
searches at the Large Hadron Collider provide strong new motivations
for supersymmetric theories with heavy superpartners.  We reconsider
focus point supersymmetry (FP SUSY), in which all squarks and sleptons
may have multi-TeV masses without introducing fine-tuning in the weak
scale with respect to variations in the fundamental SUSY-breaking
parameters.  We examine both FP SUSY and its familiar special case,
the FP region of mSUGRA/CMSSM, and show that they are beautifully
consistent with all particle, astroparticle, and cosmological data,
including Higgs boson mass limits, null results from SUSY searches,
electric dipole moments, $b \to s \gamma$, $B_s \to \mu^+ \mu^-$, the
thermal relic density of neutralinos, and dark matter searches.  The
observed deviation of the muon's anomalous magnetic moment from its
standard model value may also be explained in FP SUSY, although not in
the FP region of mSUGRA/CMSSM.  In light of recent data, we advocate
refined searches for FP SUSY and related scenarios with heavy squarks
and sleptons, and we present a simplified parameter space to aid such
analyses.
\end{abstract}


\maketitle

\section{Introduction}
\label{sec:introduction}

Since its discovery decades ago, supersymmetry (SUSY) has attracted
more attention than any other principle for physics beyond the
standard model (SM).  Of particular interest is weak-scale SUSY, which
holds the promise of providing natural resolutions to the gauge
hierarchy and dark matter problems.  For the last year, the Large
Hadron Collier has been colliding protons with protons at a
center-of-mass energy of 7 TeV.  The ATLAS and CMS experiments have
each analyzed over $1~\ifb$ of data and collected over $5~\ifb$, but
have not reported evidence for new physics~\cite{Aad:2011ib,%
Aad:2011qa,Aad:2011iu,Chatrchyan:2011ek,Chatrchyan:2011zy,Padhi:2011rs}.
These null results have excluded generic SUSY models with light
superpartners and large missing $E_T$ signatures.

Although these LHC results have disappointed the most optimistic SUSY
enthusiasts, they do not remove the possibility that weak-scale SUSY
is realized in nature.  Rather, they shift attention to supersymmetric
models that have heavier superpartners or less distinctive signatures.
The former possibility is particularly natural to consider, since
stringent constraints on flavor- and CP-violation have long motivated
heavy squarks and sleptons of the first two generations, and
experimental bounds on the Higgs boson mass have long motivated heavy
third generation squarks to raise the Higgs boson mass through large
radiative corrections.  This possibility has now received even greater
motivation from recent results from the ATLAS and CMS experiments,
which combined confine the possibility of a light Higgs boson to the
mass window $115.5~\gev < m_h < 127~\gev$, and indicate excess events
consistent with the production of Higgs bosons with masses of 126 GeV
and 124 GeV, respectively~\cite{ATLASHiggs,CMSHiggs}.  Of course, the
possibility of multi-TeV third generation squarks is generically in
tension with the requirement that SUSY resolve the gauge hierarchy
problem.

In this study, we consider focus point (FP)
SUSY~\cite{Feng:1999mn,Feng:1999zg,Feng:2000bp,Feng:2000gh} in light
of recent results.  We are motivated to consider FP SUSY for several
reasons.  First, in FP SUSY, {\em all} squarks and sleptons may be
multi-TeV without increasing the fine-tuning in the weak scale with
respect to variations in the fundamental SUSY-breaking parameters.
Naturalness is a notoriously brittle and subjective criterion, but in
this sense, FP SUSY is the unique framework that naturally
accommodates multi-TeV top and bottom squarks.  Second, many
observables, including those at colliders, in low-energy probes, and
those related to dark matter, are insensitive to the details of the
heavy scalar spectrum, since the scalars decouple.  For these
observables, FP SUSY may be viewed as an effective theory that
captures the essential features of a large class of models with heavy
superpartners. And last, a special case of FP SUSY is realized in the
FP region of minimal supergravity (mSUGRA) or the constrained minimal
supersymmetric standard model (CMSSM), heretofore referred to as the
``FP region.''  Given the amount of work devoted to this model, FP
SUSY is a practical and natural starting place for considering SUSY
models with heavy superpartners that are newly motivated by LHC data.
For other recent work on FP SUSY and the related framework of
hyperbolic branch SUSY~\cite{Chan:1997bi} motivated by recent results,
see Refs.~\cite{Asano:2011kj,Akula:2011jx}.
 
We begin in \secref{FPSUSY} by reviewing the general framework of FP
SUSY and its well-known special case, the FP region.  In
\secref{higgsmass}, we show Higgs mass predictions in mSUGRA/CMSSM,
determine the parameter space favored by Higgs mass bounds, and find
that current limits favor the FP region.  In \secref{EDMs} we show
that constraints on the electron and neutron electric dipole moments
(EDMs) are naturally satisfied in FP SUSY.  In \secref{omega}, we then
focus on the part of the FP region that has the correct neutralino
thermal relic density $\Omegachi$.  This is typically presented as a
thin strip in the $(m_0, \mgaugino)$ plane with fixed $\tan \beta$.
To allow a more comprehensive presentation of FP results, we instead
fix $m_0$ to give the correct $\Omegachi$, and present results in the
$(\tan\beta, \mgaugino)$ plane, with every point satisfying $\Omegachi
\simeq 0.23$.  In \secref{b} we present results for $b \to s \gamma$
and $B_s \to \mu^+ \mu^-$ in the $(\tan\beta, \mgaugino)$ plane, and
in \secref{directdetection} we analyze implications for dark matter
direct detection and show that FP SUSY remains consistent with current
null results.  Finally, in \secref{muon} we show that the observed
deviations of $(g-2)_{\mu}$ from SM expectations may be easily
explained in FP SUSY (but not in the FP region).  Our findings are
summarized in \secref{conclusions}.  The robustness of our numerical
analyses is discussed in the Appendix.

\section{Focus Point Supersymmetry}
\label{sec:FPSUSY}

In SUSY, the $Z$ boson mass is determined at tree-level by the
relation
\begin{equation}
\frac{1}{2} m_Z^2 = \left. - \mu^2 + \frac{m_{H_d}^2 - m_{H_u}^2 \tan^2
  \beta} {\tan^2 \beta - 1} \right|_{m_{\text{weak}}} \ ,
\label{mZ}
\end{equation}
where $\mu$ is the Higgsino mass parameter, $m_{H_{d,u}}^2$ are the
soft SUSY-breaking Higgs mass parameters, $\tan\beta \equiv \langle
H_u^0 \rangle / \langle H_d^0 \rangle$ is the ratio of Higgs boson
vacuum expectation values, and all of these are evaluated at a
renomalization group scale near $\mweak \sim 100~\gev - 1~\tev$.  For
the moderate and large values of $\tan\beta$ required by current Higgs
mass bounds, this may be simplified to
\begin{equation}
\frac{1}{2} m_Z^2 \approx \left. - \mu^2 - 
m_{H_u}^2  \right|_{m_{\text{weak}}} \ .
\label{mZ2}
\end{equation}
The weak-scale parameter $m_{H_u}^2$ depends on a set of fundamental
parameters $\{a_i \}$, typically taken to be grand unifed theory
(GUT)-scale soft SUSY-breaking parameters, such as scalar masses
$m_{\tilde{f}}$, gaugino masses $M_i$, and trilinear scalar couplings
$A_i$.  Naturalness requires that $m_Z$ not be unusually sensitive to
variations in the fundamental parameters $a_i$.  This does not
necessarily imply $a_i \sim m_Z$ for every $i$, however, because terms
involving some $a_i$ in the expression for $m_Z^2$ may be suppressed
by small numerical coefficients.

In the class of FP SUSY models studied in
Refs.~\cite{Feng:1999mn,Feng:1999zg,Feng:2000bp,Feng:2000gh}, the
fundamental GUT-scale parameters satisfy
\begin{eqnarray}
\left( m_{H_u}^2, m_{T_R}^2 , m_{(T,B){_L}}^2 \right) &=& m_0^2 \, 
(1, 1+x, 1-x) \label{lowtanb}  \\
\text{all other scalar masses} &\alt& {\cal O}(10~\tev) \\
M_i, A_i  &\alt& 1~\tev
\end{eqnarray}
for moderate $\tan\beta$, or
\begin{eqnarray}
\left( m_{H_u}^2, m_{T_R}^2, m_{(T,B){_L}}^2 , m_{B_R}^2, 
m_{H_d}^2 \right) &=& m_0^2 \, (1, 1+x, 1-x, 1+x-x', 1+x') 
\label{hightanb} \\
\text{all other scalar masses} &\alt& {\cal O}(10~\tev) \\
M_i, A_i &\alt& 1~\tev
\end{eqnarray}
for high $\tan\beta$, where the top and bottom Yukawa couplings are
comparable.  In \eqsref{lowtanb}{hightanb}, $x$ and $x'$ are arbitrary
constants, but for any values of $x$ and $x'$, the weak-scale is
insensitive to variations in $m_0$, even for multi-TeV $m_0$.  In
other words, with these GUT-scale boundary conditions, renormalization
group evolution takes $m_{H_u}^2$ to values around $m_Z^2$ at the weak
scale, almost independent of its initial GUT-scale value.  This
``focusing'' of renormalization group trajectories does not apply to
the top and bottom squark masses or, of course, to any other squark
and slepton masses.  As a result, in FP SUSY, all squarks and sleptons
may have multi-TeV masses without introducing fine-tuning in the
electroweak scale with respect to variations in the fundamental soft
SUSY-break parameters.  For an extended discussion of naturalness in
FP SUSY, see Ref.~\cite{Feng:2000bp}.

As evident from \eqsref{lowtanb}{hightanb}, the framework of FP SUSY
is quite general.  If one assumes that $x = x' = 0$, that all other
sfermion masses are also unified to the same $m_0$, that all gaugino
masses are unified, and that all $A$-parameters are unified, FP SUSY
parameter space intersects the mSUGRA/CMSSM parameter space in what is
known as the FP region.  In general, however, FP SUSY requires neither
gaugino mass nor $A$-parameter unification, and also does not
constrain scalar masses that are only weakly coupled to the Higgs
sector, such as the first and second generation squark and slepton
masses.  In much of the analysis below, we will consider the FP
region, as in many cases, it serves as an adequate representative of
general FP SUSY.  The distinction between FP SUSY and the FP region
will be relevant, however, when we discuss FP SUSY predictions for
$(g-2)_{\mu}$ in \secref{muon}.

\section{Higgs Boson Mass}
\label{sec:higgsmass}

As is well-known, current bounds from LEP2 require the Higgs boson
mass to be $m_h > 114.4~\gev$~\cite{Barate:2003sz}.  In SUSY, where
the limit $m_h \le m_Z$ applies at tree-level, large radiative
corrections from heavy top and bottom squarks are required to satisfy
this bound.  A significant phenomenological advantage of the FP SUSY
framework is that it naturally accommodates heavy third generation
squarks, and with them, relatively heavy Higgs bosons consistent with
the LEP2 bound.  Given recent Higgs boson results from the
LHC~\cite{ATLASHiggs,CMSHiggs}, it is, of course, also interesting to
investigate whether Higgs boson masses in the allowed window
$115.5~\gev < m_h < 127~\gev$ are possible, and whether masses as
large as $\sim 125~\gev$ may be naturally accommodated.

\begin{figure}
  \subfigure[\ $m_h$ and $m_{\chi}$ in GeV for $\tan\beta = 10$]{
    \includegraphics[width=.48\textwidth]{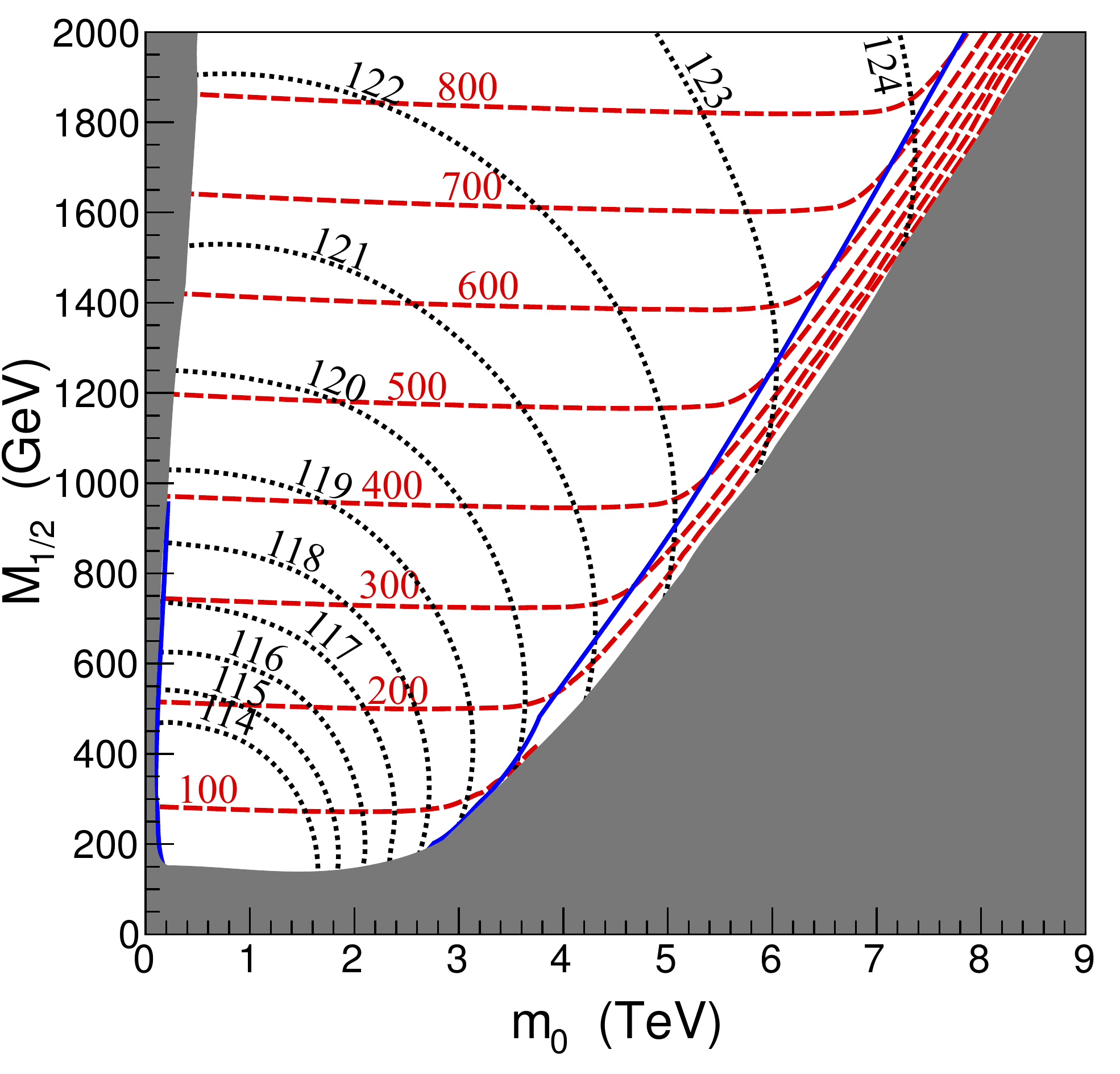}
    \label{fig:mhiggsmchi} }
  \subfigure[\ $m_h$ and $m_{\chi}$ in GeV for $\tan\beta = 50$]{
    \includegraphics[width=.48\textwidth]{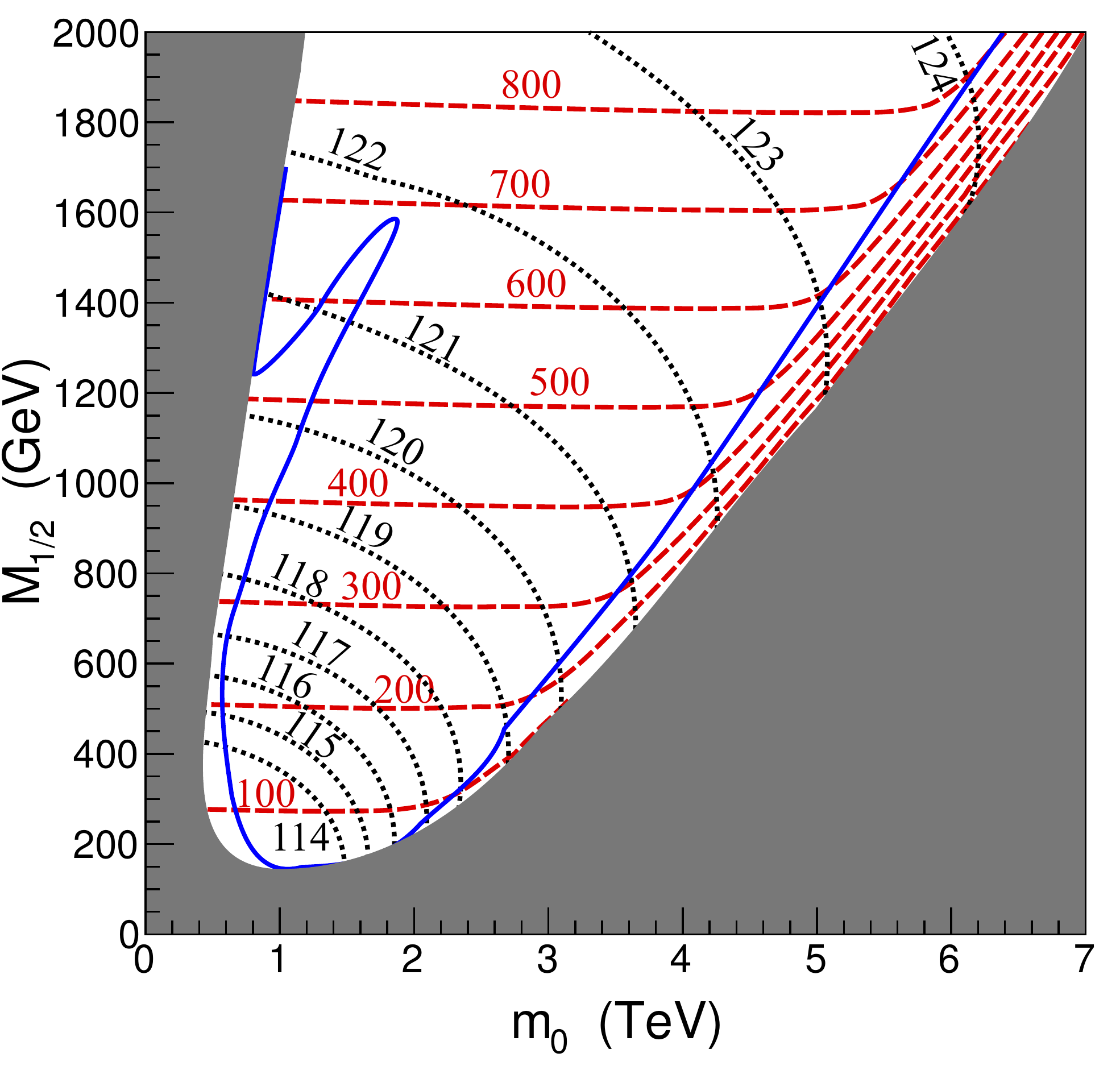}
    \label{fig:mhiggsmchitanb50} }
\caption{Contours of the light Higgs boson mass $m_h$ in black
  (dotted) and lightest neutralino mass $m_{\chi}$ in red (dashed) in
  the $(m_0, \mgaugino)$ plane for $\tan\beta = 10$ (left) and 50
  (right), $A_0 = 0$, and $\mu > 0$.  On the blue (solid) lines, the
  neutralino relic density is $\Omegachi \simeq 0.23$. }
\label{fig:mhiggsmchiall}
\end{figure}

In \figref{mhiggsmchiall}, we plot contours of constant Higgs boson
mass $m_h$ in the $(m_0, \mgaugino)$ plane of mSUGRA/CMSSM. Also shown
is the contour on which the neutralino relic density satisfies
$\Omegachi \simeq 0.23$.  Here and throughout we use
SOFTSUSY~3.1.7~\cite{Allanach:2001kg} to generate the SUSY spectrum,
and MicrOMEGAs~2.4~\cite{Belanger:2010gh} to calculate the relic
density and several other observables. In each case, we use 
a top quark mass of $m_t=173.1$ GeV and strong coupling constant
$\alpha_s(M_Z)=0.1172$.

Restricting attention to the cosmologically favored contour with
$\Omegachi \simeq 0.23$, we see that the Higgs mass bound $m_h >
114.4~\gev$ requires either $m_0 \agt 2~\tev$ (the FP region), or very
low $m_0$ and $\mgaugino \agt 500~\gev$ (the co-annihilation region).
For the parameters plotted, then, the LEP2 Higgs mass bound has
already eliminated much of the parameter space now excluded by null
results from LHC SUSY searches.  In the FP region, the Higgs boson
mass satisfies $m_h \agt 114~\gev$, and extends up to 122 GeV (124
GeV) for $\mgaugino \sim 1~\tev$ (2 TeV).  Given an estimated 2-3 GeV
uncertainty in the Higgs boson mass
calculation~\cite{Allanach:2001kg,Degrassi:2002fi,Heinemeyer:2004ms},
the FP region beautifully predicts Higgs boson masses in the currently
allowed range from 115.5 GeV to 127 GeV, and also naturally
accommodates the 124-126 GeV mass range tentatively indicated by LHC
search results.  Varying $A_0$ within the range $|A_0| \alt \tev$ can
also raise the Higgs boson mass slightly by $\sim 1~\gev$.

Contours of constant dark matter mass $m_{\chi}$ are also shown.  Note
that $m_{\chi} \sim {\cal O} (100~\gev)$, even for multi-TeV $m_0$ in
the cosmologically-favored regions.  The viable FP region contains
heavy sleptons and squarks, but potentially sub-TeV gluinos,
electroweak gauginos and Higgsinos as light as 200 GeV, and neutralino
dark matter as light as 100 GeV, even under the restrictive assumption
of gaugino mass unification.  We will return to the cosmological
implications of FP SUSY in \secref{directdetection}.

\section{Electric Dipole Moments}
\label{sec:EDMs}

FP SUSY is also motivated by constraints from EDMs.  Generic SUSY
theories with weak-scale superpartners violate low-energy flavor- and
CP-violation constraints.  Although there are well-known mechanisms to
suppress flavor violation, these do not typically suppress CP
violation.  In general, all gaugino masses, $A$-terms, and the $\mu$
parameter can possess phases that give rise to CP violation.  The most
limiting CP-violating, but flavor-conserving, observables are the EDMs
of the electron and neutron, which can arise from loop diagrams with
either left-right sfermion mixing or a gaugino-Higgsino flip within
the loop.  Even with $A \neq 0$, left-right mixing for first
generation sfermions is typically negligible, but an EDM contribution
can still arise if there is a mismatch between the phases of the
gaugino masses and the phase of $\mu$.

To examine these effects, we consider a simple extension of
mSUGRA/CMSSM where the gaugino masses and $\mu$ have general
CP-violating phases and the mismatch is parameterized as $\phiCP$.
The dominant diagrams involve left-handed sfermions and charginos
with a Wino-Higgsino mixture, leading to
contributions~\cite{Feng:2001sq}
\begin{equation}
d_f = \frac{1}{2} e \, m_f \, g_2^2 \, |M_2 \mu| \, \tan \beta \, \sin
\phiCP \, K_C ( m_{\tilde{f}_L}^2, |\mu|^2, |M_2|^2) \ ,
\end{equation}
where $K_C$ is a kinematic function~\cite{Moroi:1995yh}.  Diagrams
involving sfermions and neutralinos produce sub-dominant contributions.

The current bounds on the electron and neutron EDMs are $d_e < 1.6
\times 10^{-27}~e~\cm$~\cite{Regan:2002ta} and $d_n < 2.9 \times
10^{-26}~e~\cm$~\cite{Baker:2006ts}.  Assuming $m_u = 3~\mev$, $m_d =
5~\mev$, the naive quark model relation $d_n = (4 d_d - d_u)/3$, and
neglecting cancellations between different diagrams, we may derive
bounds on the phase mismatch $\phiCP$.  

\begin{figure}
  \subfigure[\ Upper limits on $\sin\phiCP$ for $\tan\beta = 10$]{
    \includegraphics[width=.48\textwidth]{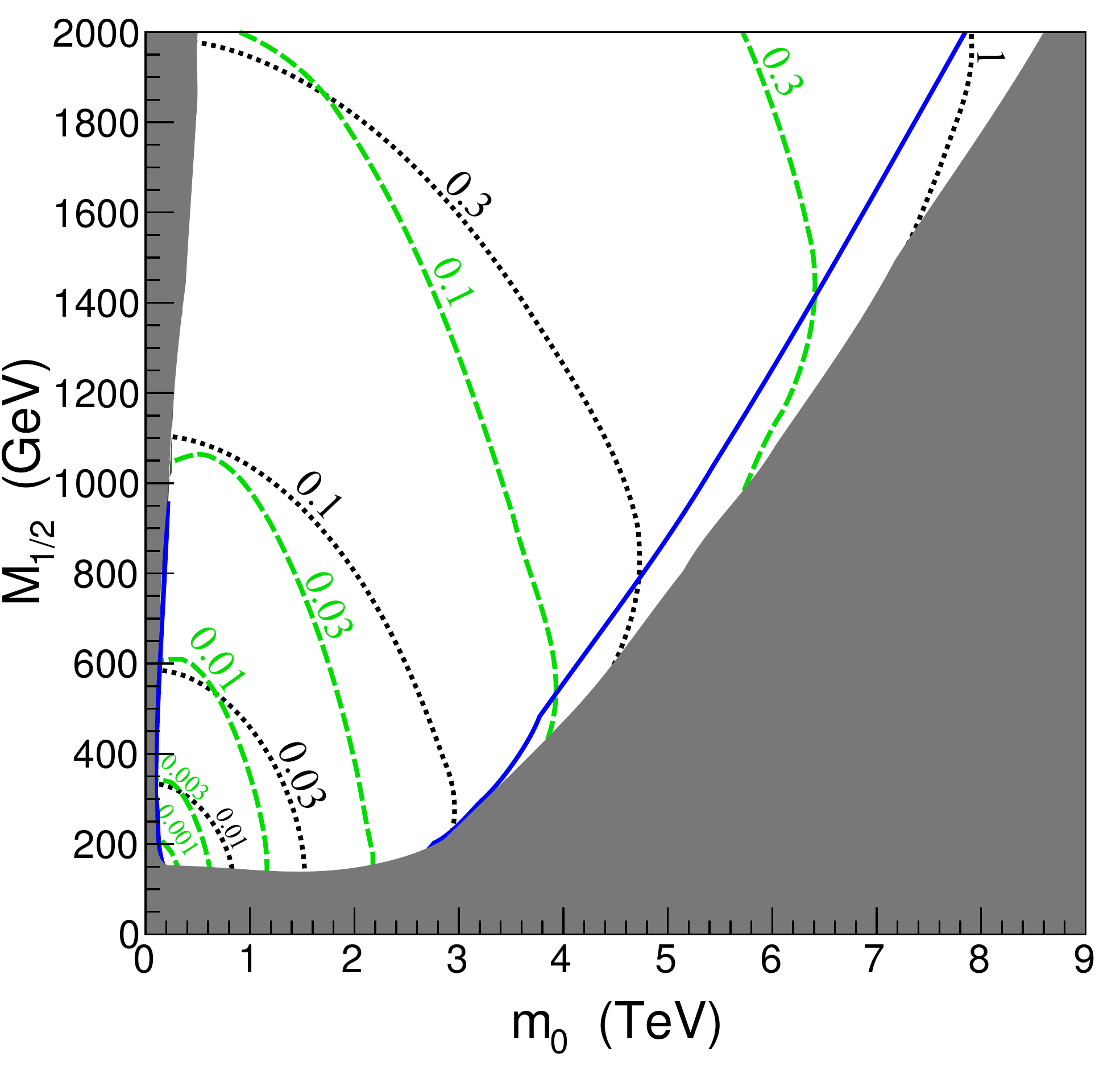}
    \label{fig:EDMtanb10} }
  \subfigure[\ Upper limits on $\sin\phiCP$ for $\tan\beta = 50$]{
    \includegraphics[width=.48\textwidth]{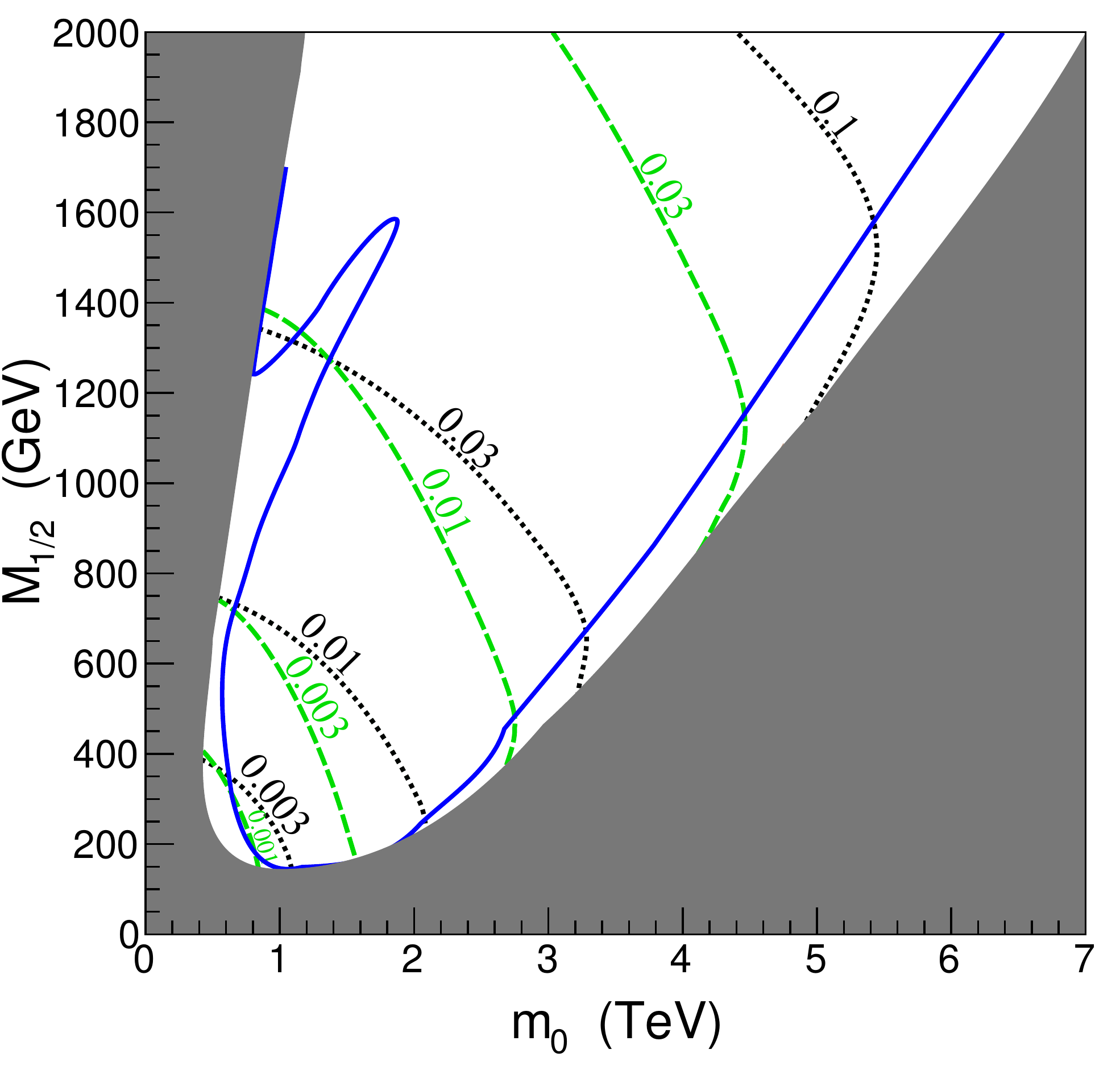}
    \label{fig:EDMtanb50} }
\caption{Upper limits on $\sin\phiCP$ from neutron EDM constraints in
  black (dotted) and electron EDM constraints in green (dashed) for
  $\tan\beta = 10$ (left) and 50 (right), $A = 0$, and $\mu > 0$. On
  the blue (solid) line, the neutralino relic density is $\Omegachi
  \simeq 0.23$. }
\label{fig:EDM}
\end{figure}

\Figref{EDM} shows the upper limits on $\sin\phiCP$ in the $(m_0,
\mgaugino)$ plane from electron and neutron EDM constraints.  In
mSUGRA, $m_{\tilde{e}_L} < m_{\tilde{u}_L} \simeq m_{\tilde{d}_L}$,
and so the electron EDM provides the stronger bound, but the neutron
EDM bound is also stringent.  From \figref{EDMtanb10}, for example, we
see that for $\tan\beta = 10$, the constraints $\Omegachi \simeq 0.23$
and $\sin\phiCP \agt 0.01$ can only be satisfied in the FP region, and
at the same time, the FP region with $\mgaugino \alt 1~\tev$ can
accommodate natural values of $\sin\phiCP \sim 0.3$.  The EDM bounds
become even stronger for large $\tan\beta$, but may be satisfied in
the FP region for $\mgaugino \sim 2~\tev$ for $\sin \phiCP \sim 0.1$.
Absent a compelling mechanism for suppressing flavor-conserving CP
violation, bounds from electron and neutron EDMs have long ago
restricted mSUGRA/CMSSM parameter space to the FP region, irrespective
of recent LHC results from SUSY and Higgs boson searches.

\section{FP SUSY with Fixed Relic Density}
\label{sec:omega}

Results for the mSUGRA/CMSSM framework are conventionally presented as
in \figsref{mhiggsmchiall}{EDM}.  In these figures, the cosmologically
desirable region with $\Omegachi \simeq 0.23$ is just a thin strip
running through the plane, and the cosmologically desirable FP region
is just a small part of that.  Given that much of the rest of the
cosmologically favored mSUGRA parameter space is now excluded,
however, as well as our focus on FP SUSY in this study, it is more
appropriate to consider a parameter space in which every point is in
the cosmologically favored part of the FP region.

For a neutralino LSP in the FP region, a significant Bino-Higgsino
mixture is required to produce a sufficiently low relic density, with
the Higgsino component increasing with $m_0$.  Thus the value of $m_0$
satisfying $\Omegachi \simeq 0.23$ for a particular set of other
parameters represents a lower bound.  If the neutralino composes only
a fraction of the relic density, $\Omegachi < 0.23$, scalar masses are
increased somewhat and the primary effect on our conclusion is a
weakening of direct detection limits.  It is also possible to
disconnect the FP effect on fine-tuning from cosmological
considerations by introducing a gravitino LSP which allows a larger
neutralino relic density to be considered; we restrict our intention
to the case of a neutralino LSP.

To satisfy the relic density constraint, we continue to consider fixed
values of $A_0$ and $\text{sign}(\mu)$, but require the neutralino to
be a thermal relic with $\Omegachi = 0.23$.  This implies a constraint
on the remaining parameters $m_0$, $\mgaugino$, and $\tan\beta$.  We
choose $\mgaugino$ and $\tan\beta$ as the free parameters, and use
$\Omegachi$ to determine $m_0$.\footnote{Alternatively, one could
  choose $\mgaugino$ and $m_0$ as the input parameters, and predict
  $\tan\beta$ \cite{Beskidt:2010va}.}  In general there are several
values of $m_0$ satisfying this condition for a particular
$(\mgaugino, \tan\beta)$ pair, arising from the co-annihilation region
at low $m_0$, the FP region at large $m_0$, and the $A$-funnel region
for moderate $m_0$ and large $\tan\beta$.  We focus on the FP region
by always choosing the largest value of $m_0$ for a given point in the
$(\mgaugino, \tan\beta)$ plane.

\begin{figure}[tb]
  \subfigure[\ $m_0$ (\tev)]{
    \includegraphics[width=.48\textwidth]{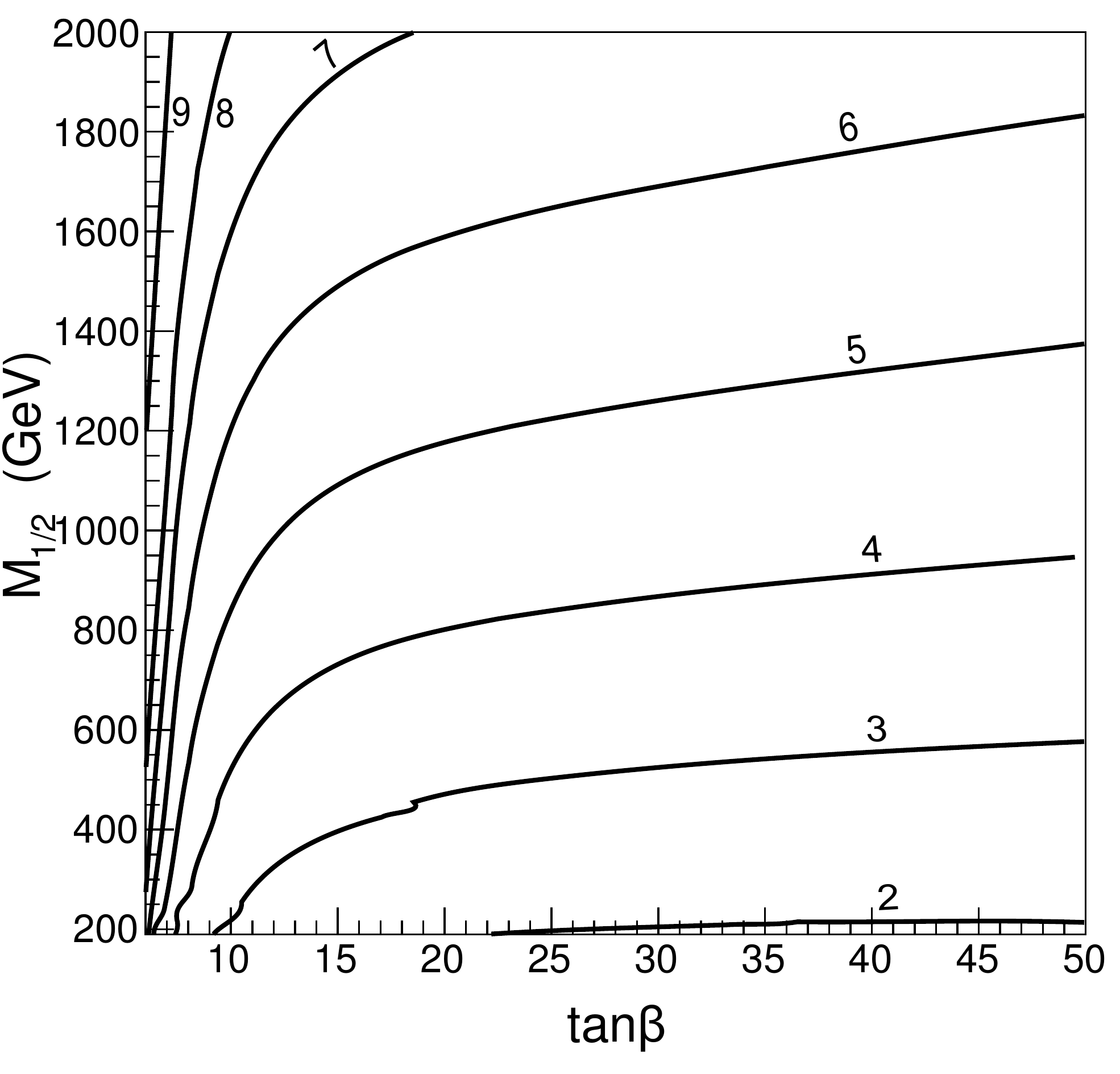}
    \label{fig:m0} }
  \subfigure[\ $\mu$ (\gev)]{
    \includegraphics[width=.48\textwidth]{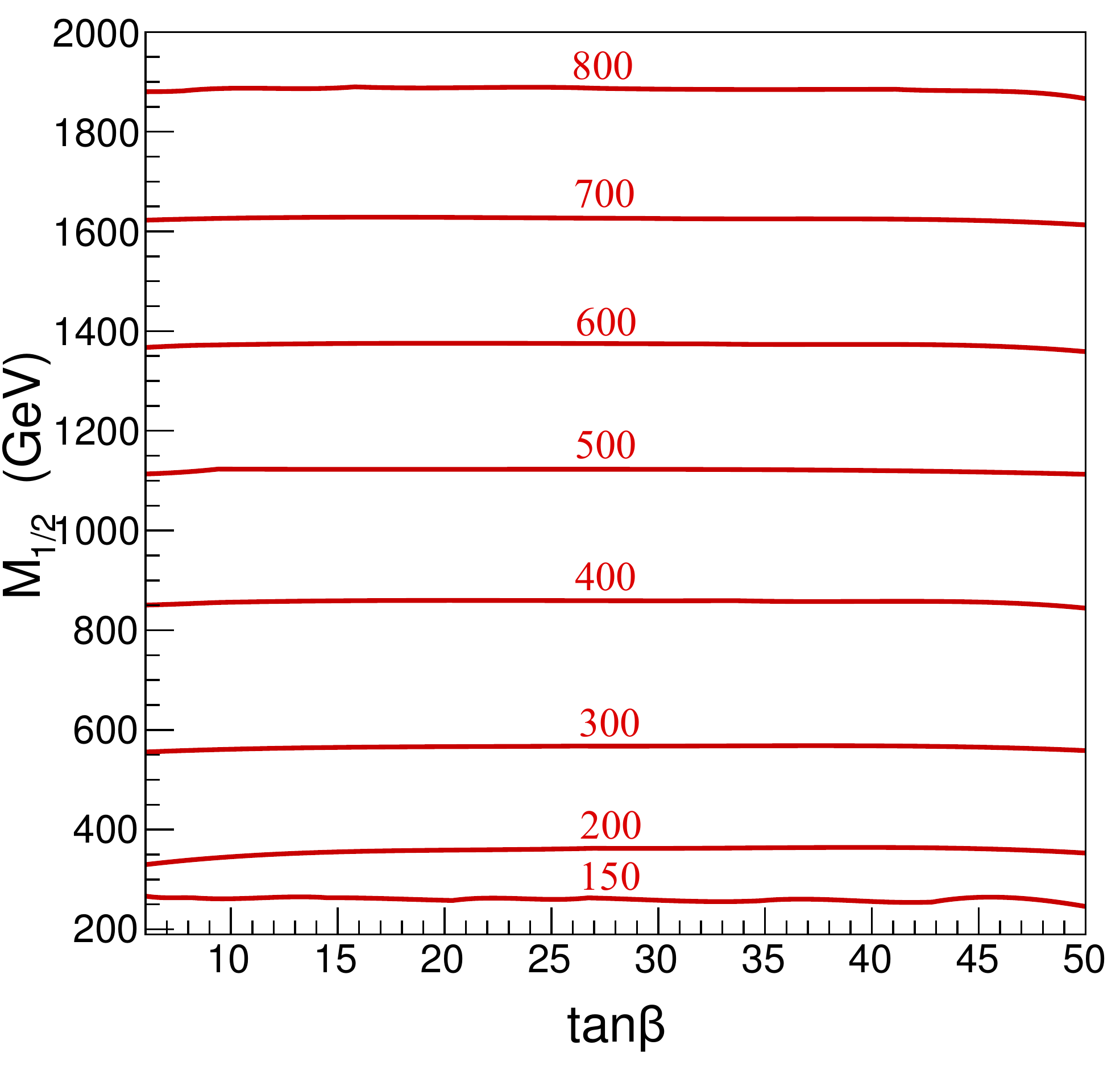}
    \label{fig:mu} }
\caption{Contours of (a) $m_0$ (in TeV) and (b) $\mu$ (in GeV) in the
$(\tan\beta, \mgaugino)$ plane.  Every point in the parameter space is
in the FP region and satisfies $\Omegachi \simeq 0.23$, $A_0 = 0$, and
$\mu > 0$.}
\label{fig:m0mu}
\end{figure}

\begin{figure}[tb]
  \subfigure[\ $m_h$ (\gev)]{
    \includegraphics[width=.48\textwidth]{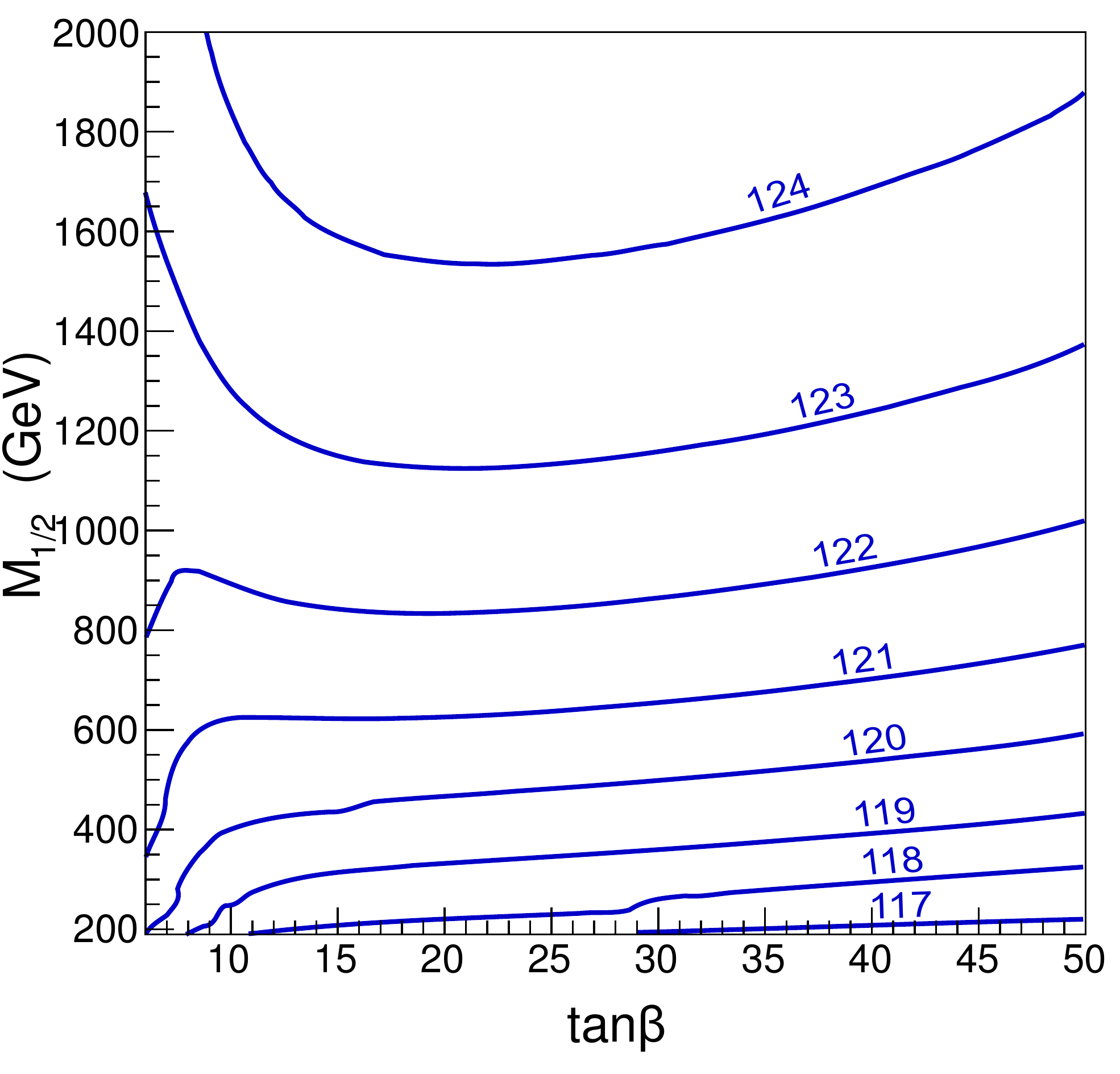}
    \label{fig:mhiggs} }
  \subfigure[\ $m_{\chi}$ (\gev)]{
    \includegraphics[width=.48\textwidth]{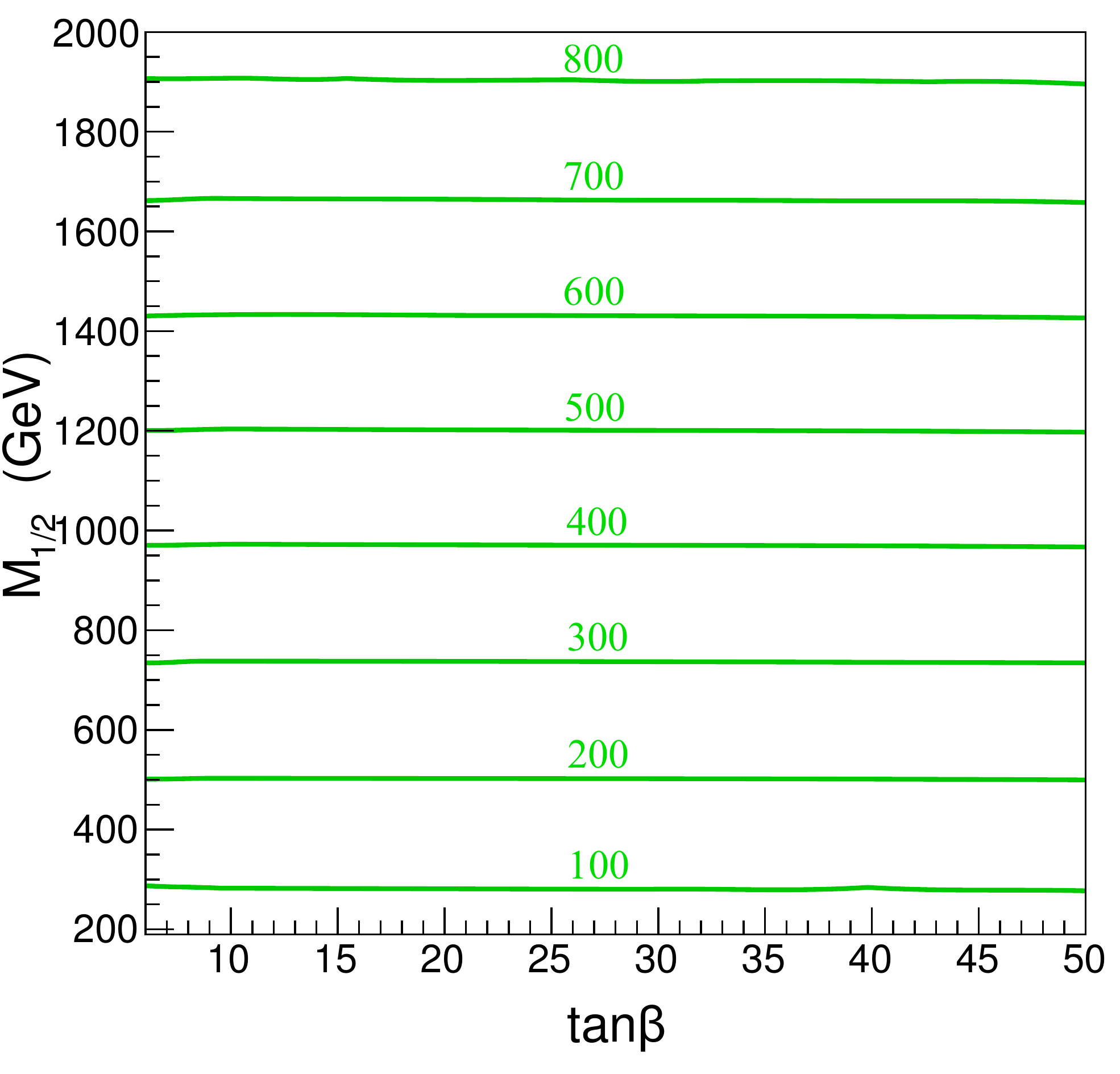}
    \label{fig:mchi} }
\caption{Contours of $m_h$ (left) and $m_\chi$ (right) in the
  $(\tan\beta, \mgaugino)$ plane.  Every point in the parameter space
  is in the FP region and satisfies $\Omegachi \simeq 0.23$, $A_0 =
  0$, and $\mu > 0$.}
\label{fig:mhiggsmchitanbeta}
\end{figure}

In \figref{m0mu}, we show contours of constant $m_0$ and $\mu$ in the
$(\tan\beta, \mgaugino)$ parameter space defined above, where every
point has $\Omegachi \simeq 0.23$, $A_0=0$ and $\mu > 0$.  In
\figref{m0}, we see that $m_0$ increases as $\mgaugino$ increases and
decrease as $\tan \beta$ increases. In the FP region the large mass of
the sfermions makes them nearly decoupled for the relic density
calculation.  The correct value of $m_0$ is instead solely determined
by its impact on the Higgs potential, which sets $|\mu|$, and which in
turn determines the correct Higgsino-Bino mixture to produce
$\Omegachi = 0.23$.\footnote{The determination of $m_0$ in \figref{m0}
  is sensitive to the value of the top mass (see, e.g.,
  Ref.~\cite{Ellis:2001zk}), and varies somewhat for different MSSM
  spectrum generation programs.  The determination of $\mu$ shown in
  \figref{mu}, however, is preformed directly from a fit to the
  measured relic density and is thus robust and independent of the
  value for the top mass or the spectrum generator
  used~\cite{Baer:2005ky}. For more details, see the Appendix.}  In
\figref{mu}, we see that $\mu$ grows with increasing $\mgaugino$, but
is nearly independent of $\tan\beta$, given the subdominance of terms
involving $\tan\beta$ in the neutralino mass matrix.

In \figref{mhiggsmchitanbeta}, we plot contours of $m_h$ and $m_\chi$
in the same $(\tan \beta, \mgaugino)$ parameter space. The large value
of $m_{\tilde{t}}$ in the FP region raises the Higgs mass well above
the LEP2 bound of $114.4~\gev$, and is confined to the currently
allowed range of $115.5~\gev < m_h < 127~\gev$.  As one moves to
smaller values of $\tan\beta$, $m_h$ increases even though its
tree-level value drops, because of the enhancement of the loop-level
contribution from increasing $m_0$.  The neutralino mass contours
satisfy $m_{\chi} \approx M_1 \simeq 0.4 \mgaugino$, since the
neutralino is primarily Bino-like, although there is an increasingly
significant Higgsino component as $\mgaugino$ increases.  As with the
$M_1$ and $\mu$ contours, the $m_\chi$ contours are also nearly
independent of $\tan\beta$.

\section{Rare B Processes}
\label{sec:b}

Rare decays are often used to constrain new physics scenarios, and in
particular, the decays $\bar{B} \to X_s \gamma$ and $B_s \to \mu^+
\mu^-$ are well-known probes of new physics.  The measured value of
$B(\bar{B} \to X_s \gamma)$ is $(3.55 \pm 0.33) \times
10^{-4}$~\cite{Asner:2010qj}, consistent with the SM value of
$(3.15\pm 0.23)\times 10^{-4}$~\cite{Misiak:2006zs,Misiak:2006ab}.
The value of $B(B_s \to \mu^+ \mu^-)$ has been the subject of recent
interest, with a CDF analysis reporting a value of $1.8^{+1.1}_{-0.9}
\times 10^{-8}$, and claiming $4.6 \times 10^{-9} < B( B_s \to \mu^+
\mu^-) < 3.9 \times 10^{-8}$ at 90\%
C.L.~\cite{Aaltonen:2011fi,Kuhr:2011kr}.  Meanwhile, CMS and LHCb
analyses produced only upper limits at 90\% C.L. of $1.9 \times
10^{-8}$~\cite{Chatrchyan:2011kr} and $5.6 \times
10^{-8}$~\cite{Serrano:2011px}, respectively, and $1.08 \times
10^{-8}$~\cite{CMSLHCbBsmumu} from a combined analysis using 2010 LHCb
data~\cite{Aaij:2011rj}.  The SM value is $(3.19\pm 0.19)\times
10^{-9}$~\cite{Buras:2003td,Gamiz:2009ku}, consistent with the LHC
bounds and marginally inconsistent with the CDF analysis.

\begin{figure}
  \subfigure[\ $\Delta B(\bar{B} \to X_s \gamma)$]{
    \includegraphics[width=.48\textwidth]{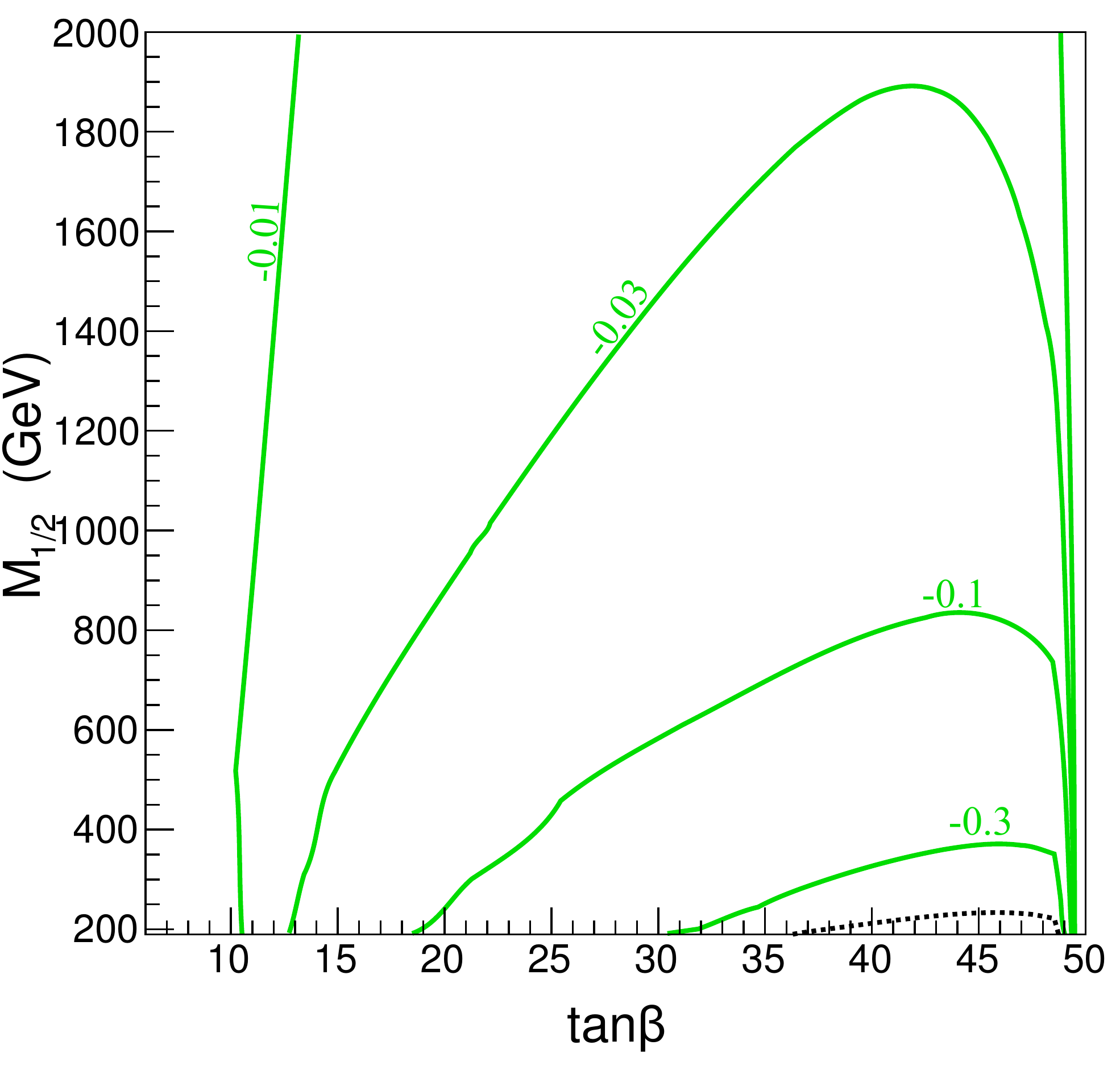}
    \label{fig:bsgamma} }
  \subfigure[\ $\Delta B(B_s \to \mu^+ \mu^-)$]{
    \includegraphics[width=.48\textwidth]{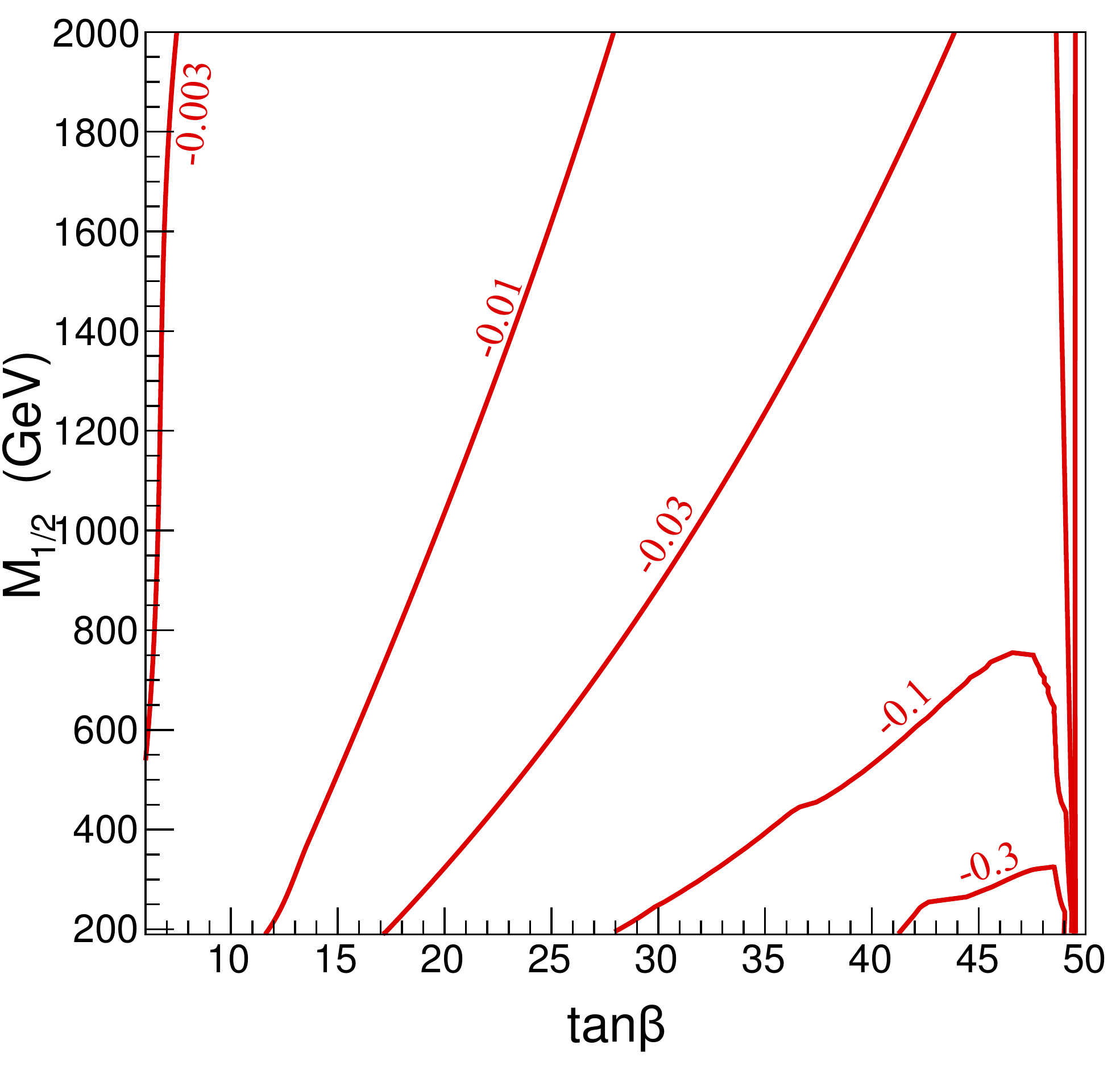}
    \label{fig:bsmumu} }
\caption{Contours of $\Delta B(b \to s \gamma$) in units of $10^{-4}$
(left) and $\Delta B(B_s \to \mu^+ \mu^-$) in units of $10^{-8}$
(right) due to SUSY in the $(\tan\beta, \mgaugino)$ plane.  Every
point in the parameter space is in the FP region and satisfies
$\Omegachi \simeq 0.23$, $A_0 = 0$, and $\mu > 0$.}
\label{fig:bdecays}
\end{figure}

\Figref{bdecays} shows the contributions to $\bar{B} \to X_s \gamma$
and $B_s \to \mu^+ \mu^-$ from supersymmetric particles.  For both
observables, the primary supersymmetric contributions arise from loop
diagrams involving either charginos or charged Higgs bosons.  For
$\bar{B} \to X_s \gamma$ the former produces a suppression in the
decay for $\mu > 0$ and an enhancement for $\mu < 0$ and the latter an
enhancement for either sign of $\mu$.  For $B_s \to \mu^+ \mu^-$ the
chargino contribution is negative and charged Higgs contribution
positive for either sign of $\mu$.  Within the FP region the chargino
diagram dominates.  For $\bar{B} \to X_s \gamma$, this puts the
supersymmetric result in greater tension with experiment than the SM
result for $\mu > 0$, though only significantly so at low $\mgaugino$
and large $\tan\beta$ --- the $2\sigma$ discrepancy line is plotted in
\Figref{bsgamma}.  For $\mu < 0$ the contribution is positive and
within $2\sigma$ of the observed result for the entire $(\mgaugino,
\tan\beta)$ plane.  For $B_s \to \mu^+ \mu^-$, the supersymmetric
contribution in the FP region does not significantly alter the SM
prediction, at least relative to current experimental uncertainties.

\section{Direct Detection of Dark Matter}
\label{sec:directdetection}

In the cosmologically-favored region of the FP region, neutralinos
make up the dark matter.  These regions of parameter space are then
constrained by null results from dark matter searches.  In particular,
null results from direct detection searches that constrain the
spin-independent $\chi$-nucleon cross section $\sigma^p$ have been
advanced as significant constraints on FP
SUSY~\cite{Farina:2011bh,Buchmueller:2011ki,Bertone:2011nj}.

In the absence of large left-right mixing, the dominant contributions
to both neutralino annihilation and spin-independent scattering are
dependent on the ``Higgsino-ness'' of the lightest neutralino, defined
as
\begin{equation}
\ah \equiv \sqrt{|\ahu|^2 + |\ahd|^2} \ ,
\label{ahdef}
\end{equation}
where the neutralino eigenstate is
\begin{equation}
\Neut = \ab \Bino + \aw \Wino + \ahu \Higgsino_u + \ahd \Higgsino_d \ .
\end{equation}

\begin{figure}
  \subfigure[\ Higgsino content (\ref{ahdef}) for various relic densities]{
    \includegraphics[width=.48\textwidth]{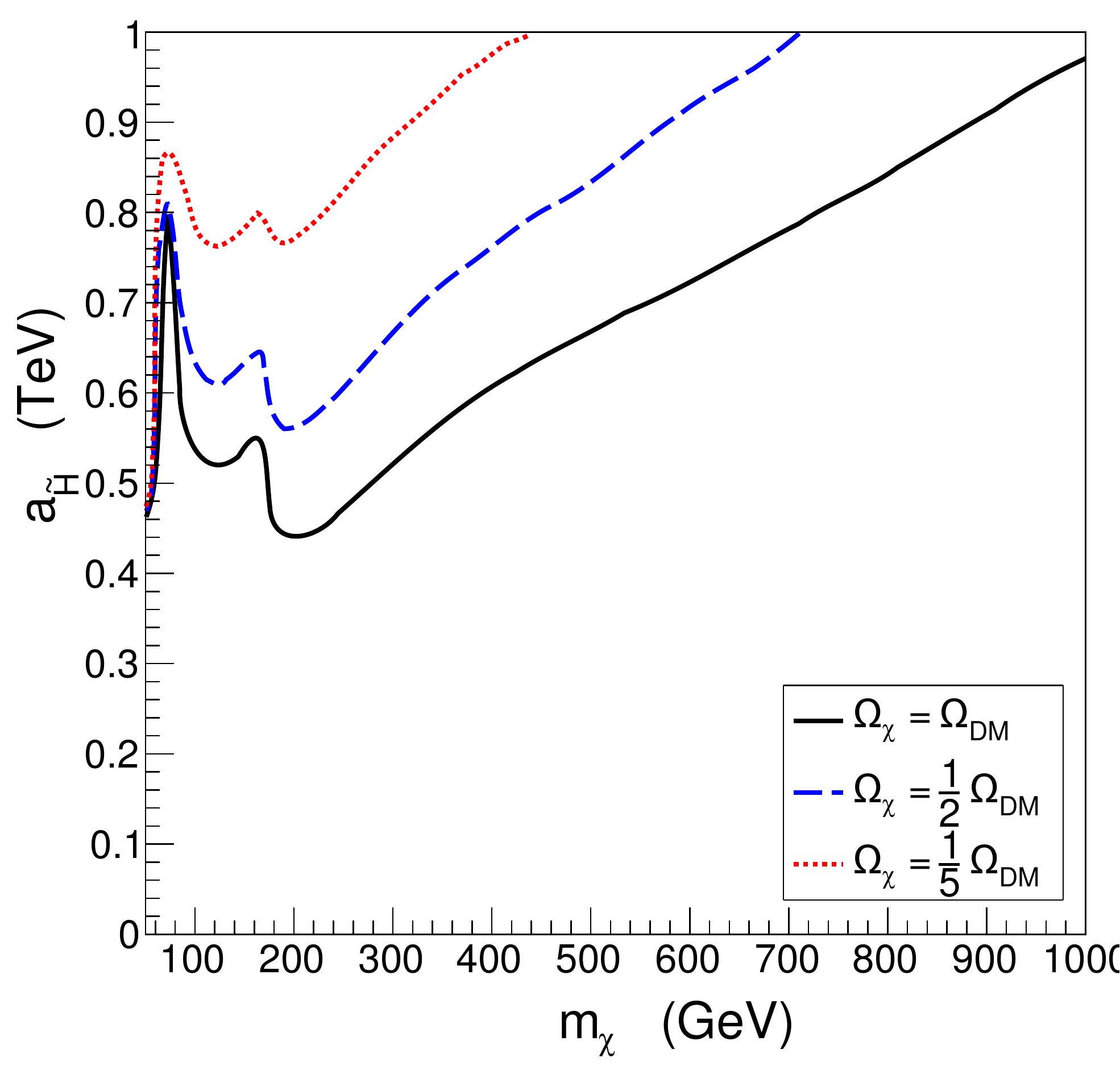}
    \label{fig:ah} }
  \subfigure[\ Dependence of $\sigma^p$ on the strange quark form
    factor $f_s$]{ \includegraphics[width=.48\textwidth]{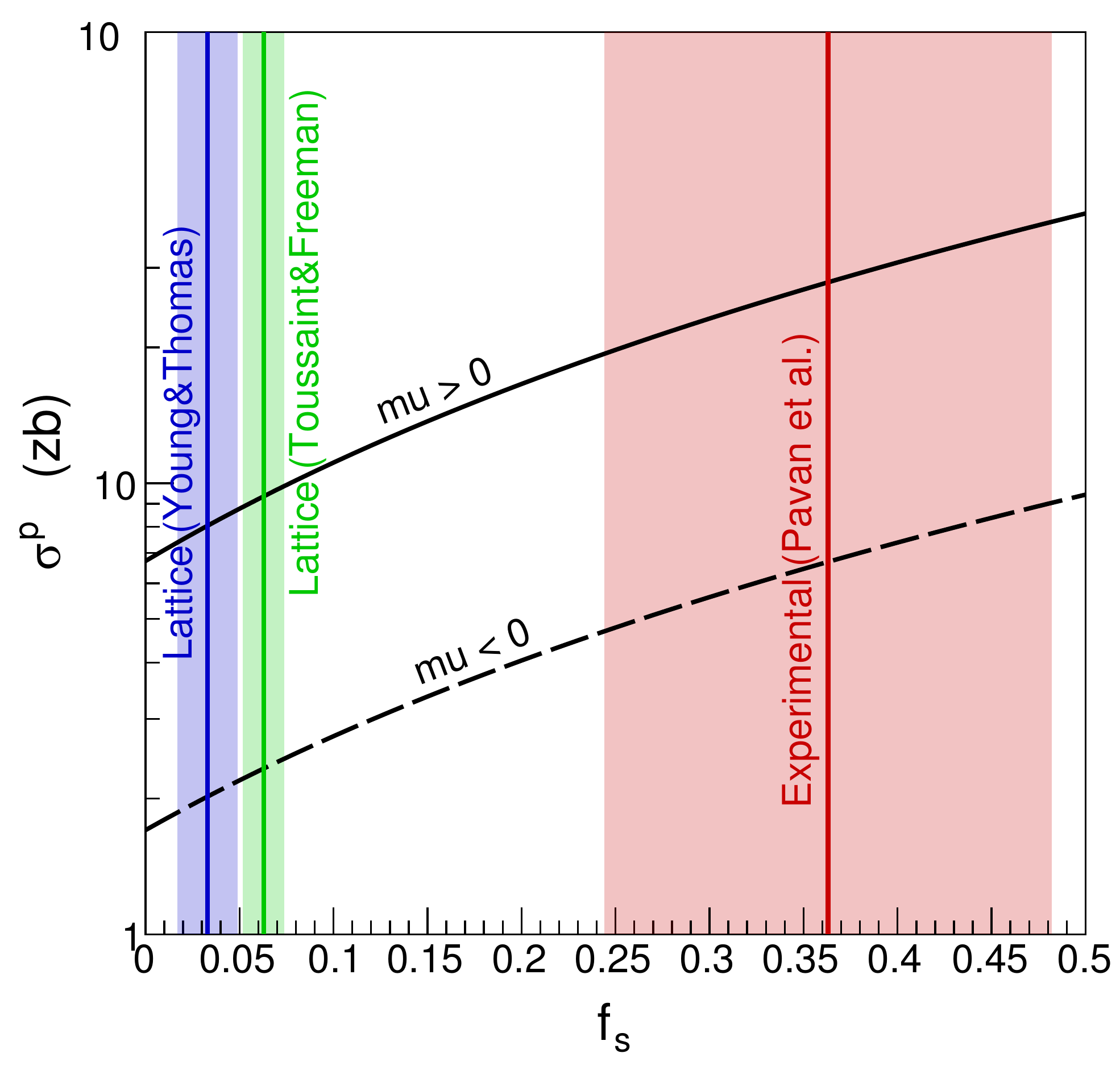}
    \label{fig:fs} }
\caption{\textit{Left}: Higgsino-ness $a_{\tilde{H}}$ for neutralinos
  in the FP region consistent with various relic densities.  The 
  right-most point on each curve corresponds to $\mgaugino =
  1~\tev$.  \textit{Right}: The spin-independent $\chi$-nucleon cross
  section $\sigmaSI_p$ as a function of $f_s$ for a model in the FP
  region with $(m_0, \mgaugino) = (3~\tev, 550~\gev)$.  The shaded
  regions indicate the $1\sigma$ uncertainties on the various $f_s$
  determinations.  In both plots, $\tan\beta = 10$, $A_0= 0$, and $\mu
  > 0$.}
\label{fig:ahfs}
\end{figure}

\Figref{ah} shows the dependence of $\ah$ on $\mchi$ in the FP region.
The Higgsino-ness generically increases with $\mchi$ to offset the
suppression in annihilation from the lowered cross section.  However,
it decreases when new annihilation channels open at $\mchi \sim m_W,
m_Z$ and $\mchi \sim m_t$.  \Figref{ah} also shows curves in which the
neutralino makes up only a fraction of the relic density --- for lower
relic densities, $\ah$ increases to enhance the annihilation rate.
The curves are generated by varying $\mgaugino$ up to $1~\tev$, for
fixed $\tan\beta = 10$, $A_0= 0$, and $\mu > 0$.

To determine the spin-independent $\chi$-nucleon cross section
$\sigma^p$, the contributions of the couplings to each individual
quark must be considered.  The individual couplings must be weighted
according to the scalar quark form factors $f_q^N$, typically
parameterized as
\begin{equation}
\left\langle N \left| m_q \bar{\psi}_q \psi_q \right| N \right\rangle
= f_q^N M_N \ .
\end{equation}
The parameters $f_{u,d}^N$ are reasonably well known, and the heavy
quark contributions are set by loop contributions using the gluon form
factor.  However, the value of $f_s^N$ is less certain, given
discrepancies between current experimental and lattice results, and
this is a well-known source of uncertainty for direct detection
predictions~\cite{Ellis:2008hf,Giedt:2009mr,Buchmueller:2011ki}.  The
experimental determination combines a derivation of the pion-nucleon
sigma term from meson scattering data~\cite{Pavan:2001wz} combined
with a number of chiral perturbation theory
results~\cite{Borasoy:1996bx,Gasser:1990ap,Bernard:1996nu}, giving
\begin{equation}
f_s = f_s^n = f_s^p \sim 0.36\ .
\end{equation}
More recent calculations support older determinations of the
pion-nucleon sigma term~\cite{Alarcon:2011zs}.  For this value of
$f_s$, the other form factors are all much smaller, $f_{q \neq s}^N
\alt 0.05$, and so the strange quark contribution dominates the direct
detection cross section~\cite{Ellis:2008hf}.  However, two recent
lattice studies have found much smaller values for
$f_s$~\cite{Young:2009zb,Freeman:2009pu}, with an average of $f_s
\approx 0.05$.  For this value of $f_s$, the strange quark
contribution is much closer to that of the other quark
flavors~\cite{Giedt:2009mr,Thomas:2011cg}.

\Figref{fs} shows the dependence of $\sigma^p$ on $f_s$ for both
positive and negative $\mu$ in the FP region.  The value of $\sigma^p$
varies by a factor of $\sim 3$ between the experimental and lattice
determinations of $f_s$, which has significant implications for direct
detections bounds.  The scattering cross section may also be
suppressed if $\mu < 0$.  This possibility is often ignored in studies
that assume $\mu > 0$ to reduce the discrepancy in $(g-2)_\mu$ between
the SM and experimental data.

\begin{figure}
  \subfigure[\ $f_s = 0.05,~\mu > 0$]{
    \includegraphics[width=.48\textwidth]{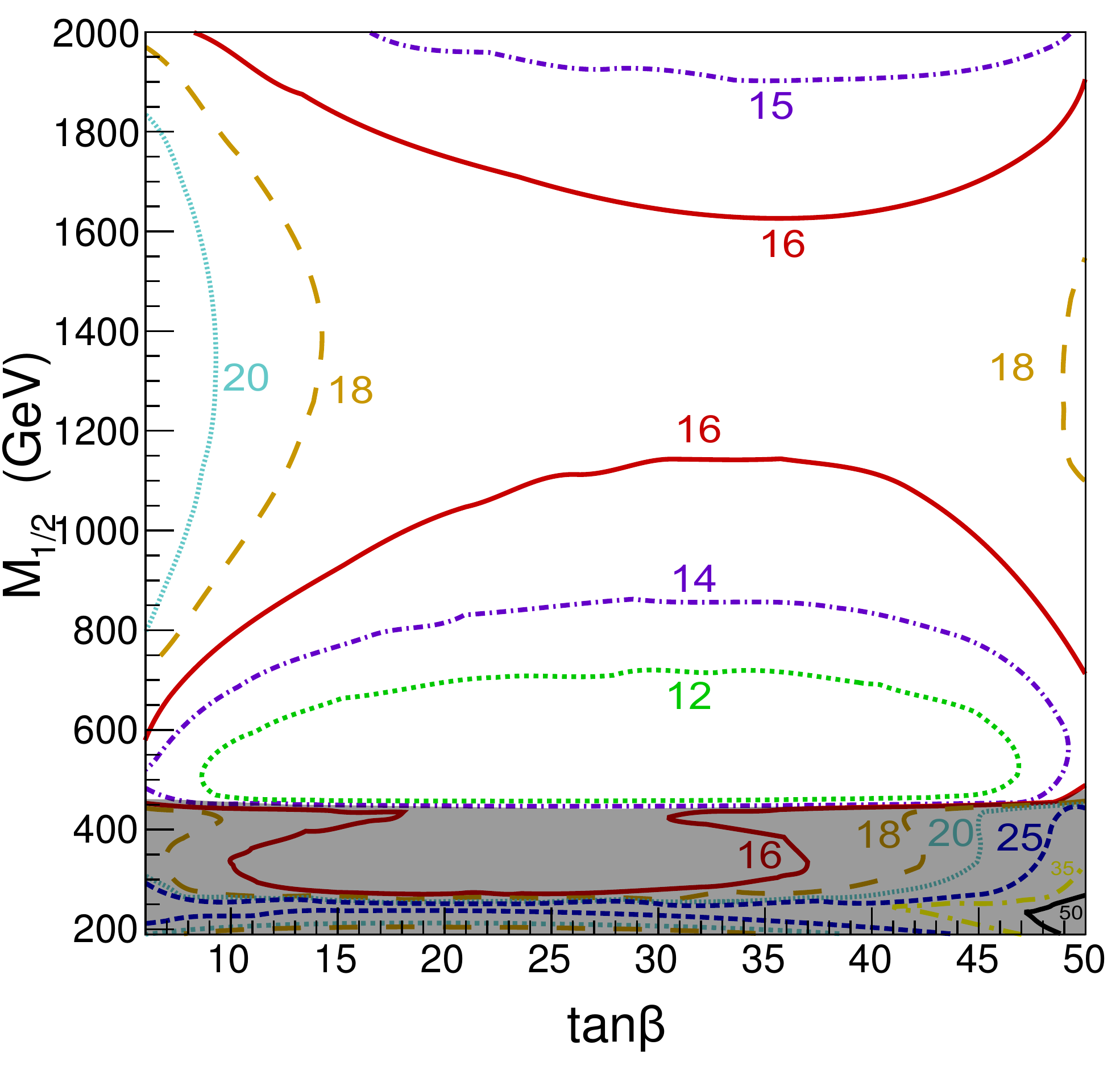}
    \label{fig:sigma_lowfs} }
  \subfigure[\ $f_s = 0.36,~\mu > 0$]{
    \includegraphics[width=.48\textwidth]{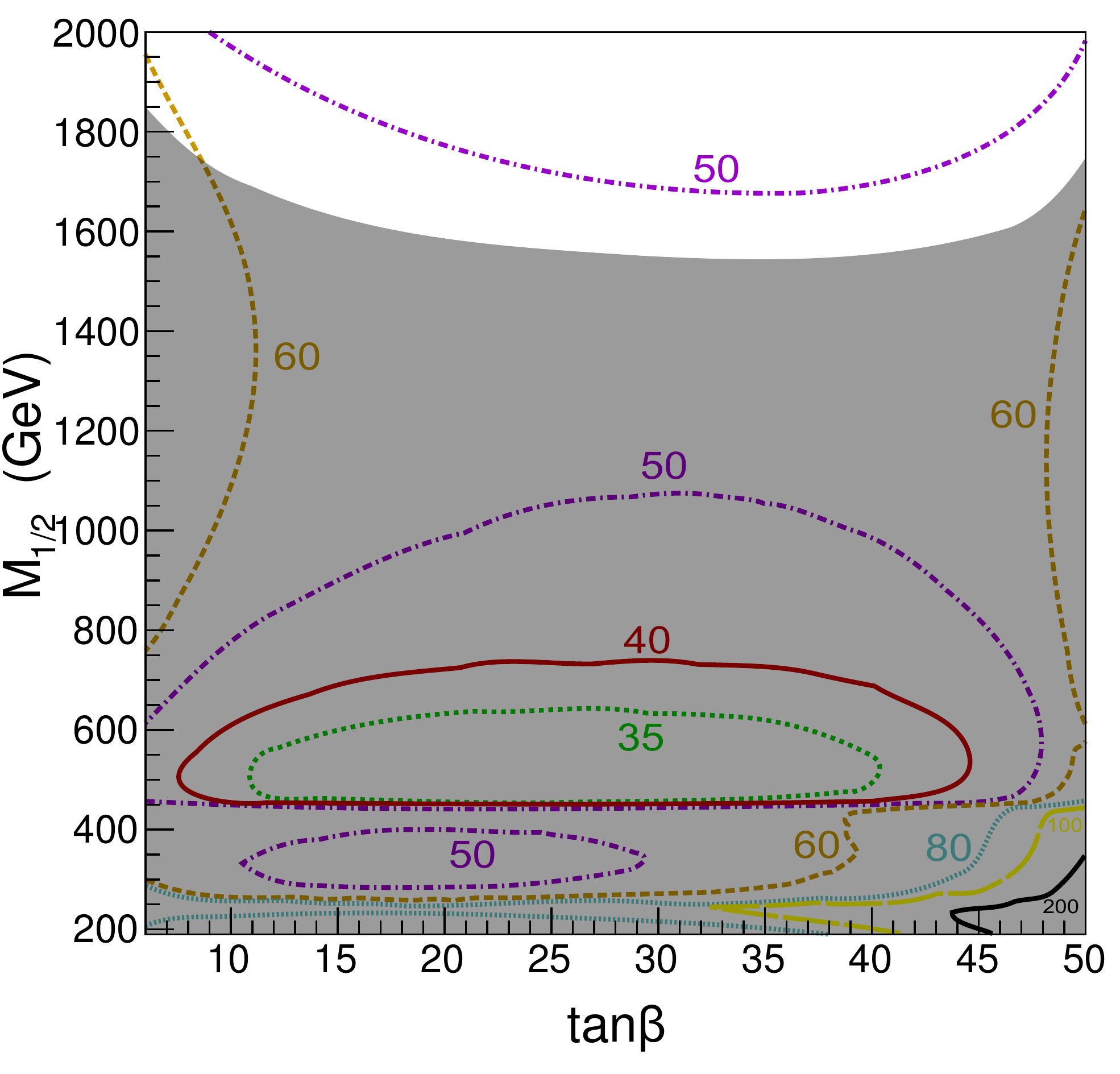}
    \label{fig:sigma_highfs} }
  \subfigure[\ $f_s = 0.05,~\mu < 0$]{
    \includegraphics[width=.48\textwidth]{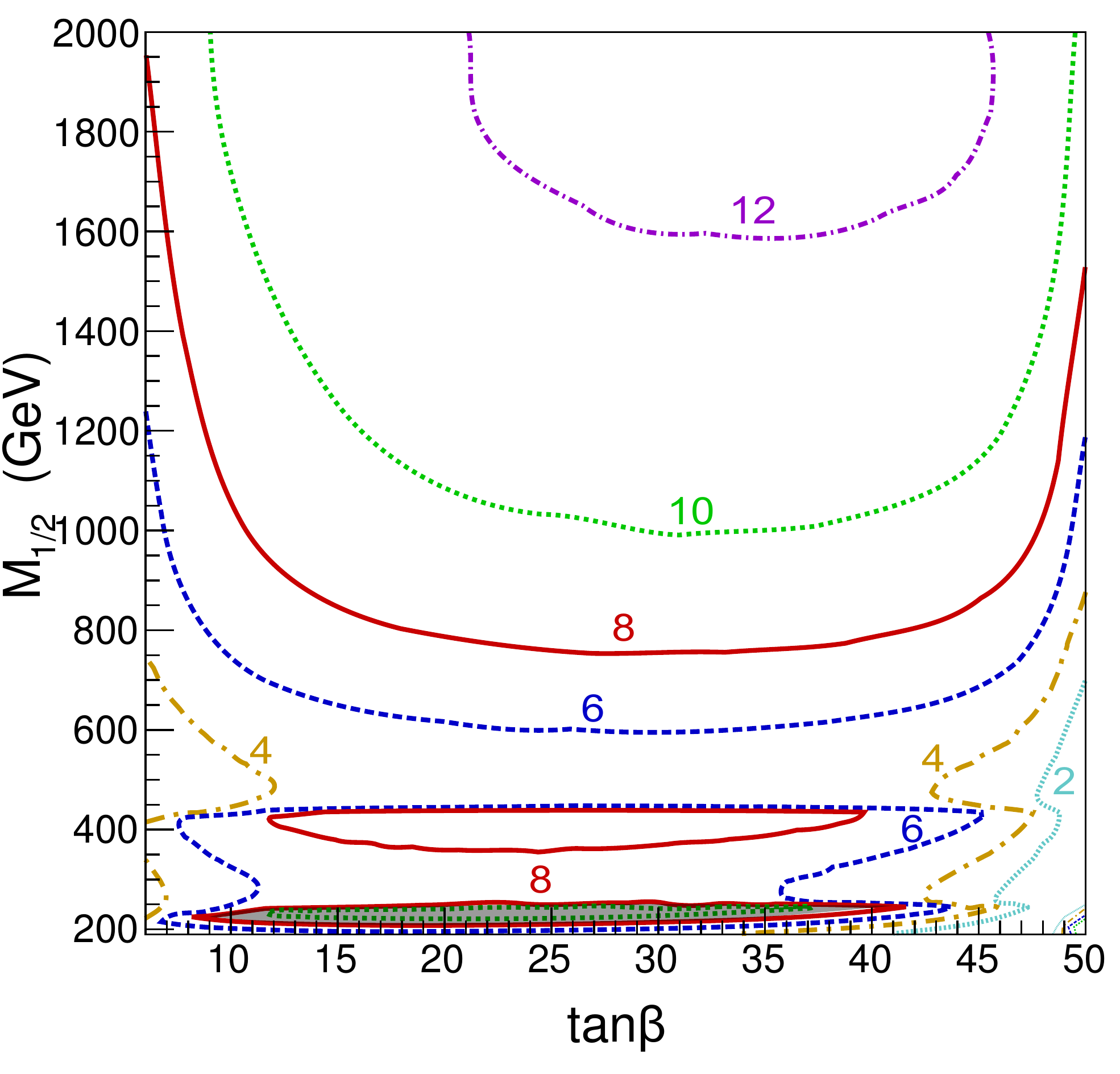}
    \label{fig:sigma_lowfs_muneg} }
  \subfigure[\ $f_s = 0.36,~\mu < 0$]{
    \includegraphics[width=.48\textwidth]{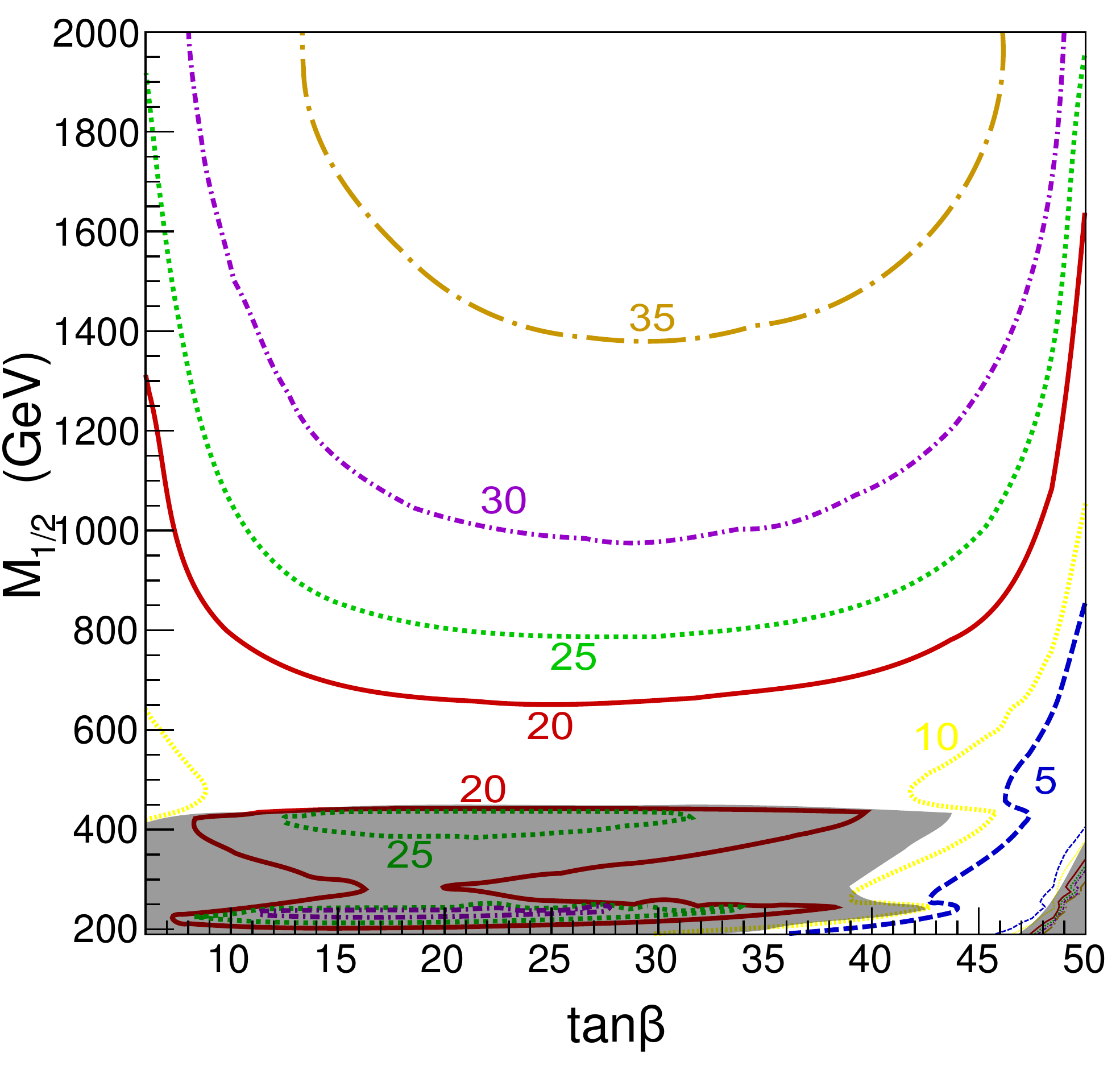}
    \label{fig:sigma_highfs_muneg} }
\caption{Contours of $\sigma^p$ in zeptobarns for $\mu > 0$ (top) and
  $\mu < 0$ (bottom), with $f_s = 0.05$ (left) and $f_s = 0.36$
  (right).  In each panel, the shaded region is excluded by
  XENON100~\cite{Aprile:2011hi}.}
\label{fig:sigmas}
\end{figure}

\Figref{sigmas} shows contours of $\sigma^p$ for positive and negative
$\mu$ and $f_s$ consistent with experimental and lattice results.  The
general factor of $\sim 3$ due to different values of $f_s$ is once
again apparent.  The cross section $\sigma^p$ for $\mu < 0$ also shows
a general suppression relative to that for $\mu > 0$, though the
suppression varies significantly with both mass scale and $\tan\beta$.
For $\mu > 0$ there is a general enhancement in $\sigma^p$ at low
$\tan\beta$ due to the coupling to the light Higgs, and at high
$\tan\beta$ due to a reduction in the masses of the heavy Higgs
bosons.  These effects are also present for $\mu < 0$, but instead
produce a suppression.

\Figref{sigmas} also shows the regions of parameter space excluded by
XENON100~\cite{Aprile:2011hi}.  For $\mu > 0, f_s = 0.05$ in
\figref{sigma_lowfs}, $\mgaugino < 450~\gev$ is excluded for all
$\tan\beta$.  The case of $f_s = 0.36$ in \figref{sigma_highfs} is
markedly different, with exclusion up to $\mgaugino \approx 1.6~\tev$
for all $\tan\beta$ and larger $\mgaugino$ for low and high
$\tan\beta$.  The same trend carries over to $\mu < 0$ --- in
\figref{sigma_lowfs_muneg} the exclusion is limited to a small range
of $\mgaugino$ at moderate $\tan\beta$ where $\ah$ peaks, and to a
small region at high $\tan\beta$ where scattering is dominated by the
heavy-Higgs boson mediated process.  The exclusion in
\figref{sigma_highfs_muneg} is greater due to larger $f_s$ but still
reduced compared to the $\mu > 0$ case.

In summary, we find that FP SUSY is far from excluded by current
direct detection bounds.  For large $f_s \sim 0.36$ and $\mu>0$,
significant portions of the parameter space are excluded, requiring
$\mgaugino \agt 1.6~\tev$ to survive.  For the smaller values of $f_s$
favored by lattice results or $\mu<0$, a much larger portion of the
parameter space is viable, including regions with gaugino masses as
low as $\mgaugino \sim 250~\gev$.  At the same time, it is, of course,
interesting that the direct detection bounds are within factors of a
few from probing all of FP SUSY. To the extent that LHC SUSY and Higgs
boson results motivate SUSY with heavy squarks and sleptons, they also
motivate direct detection experiments that are approaching
sensitivities to zeptobarn spin-independent cross sections in the near
future.

\section{The Anomalous Magnetic Moment of the Muon}
\label{sec:muon}

The well-known $\sim 3\sigma$ discrepancy between the experimental and
SM values in the anomalous magnetic moment of the
muon~\cite{Bennett:2006fi,Jegerlehner:2009ry,Davier:2010nc} is
currently among the most compelling pieces of evidence for new
physics.  The supersymmetric contribution is given by
$\tilde{\mu}-\chi^0$ and $\tilde{\nu}_\mu-\chi^\pm$ loop diagrams.
The $(g-2)_\mu$ discrepancy has two robust implications for SUSY ---
it is the primary result motivating relatively light superpartners,
and it favors $\mu > 0$.

The large sfermion masses in the FP region produce too small a value
for $\Delta (g-2)_\mu^{\text{SUSY}}$ to explain the observed
discrepancy of $(2.9\pm 0.9) \times
10^{-9}$~\cite{Jegerlehner:2009ry}.  \Figref{gmuon} shows the value of
$\Delta (g-2)_\mu^{\text{SUSY}}$ in the FP region parameter space.
The largest value attained is $\Delta (g-2)_\mu^{\text{SUSY}} \approx
0.5 \times 10^{-9}$, insufficient to produce even $2\sigma$ agreement
with the experimental result.

\begin{figure}
    \includegraphics[width=.48\textwidth]{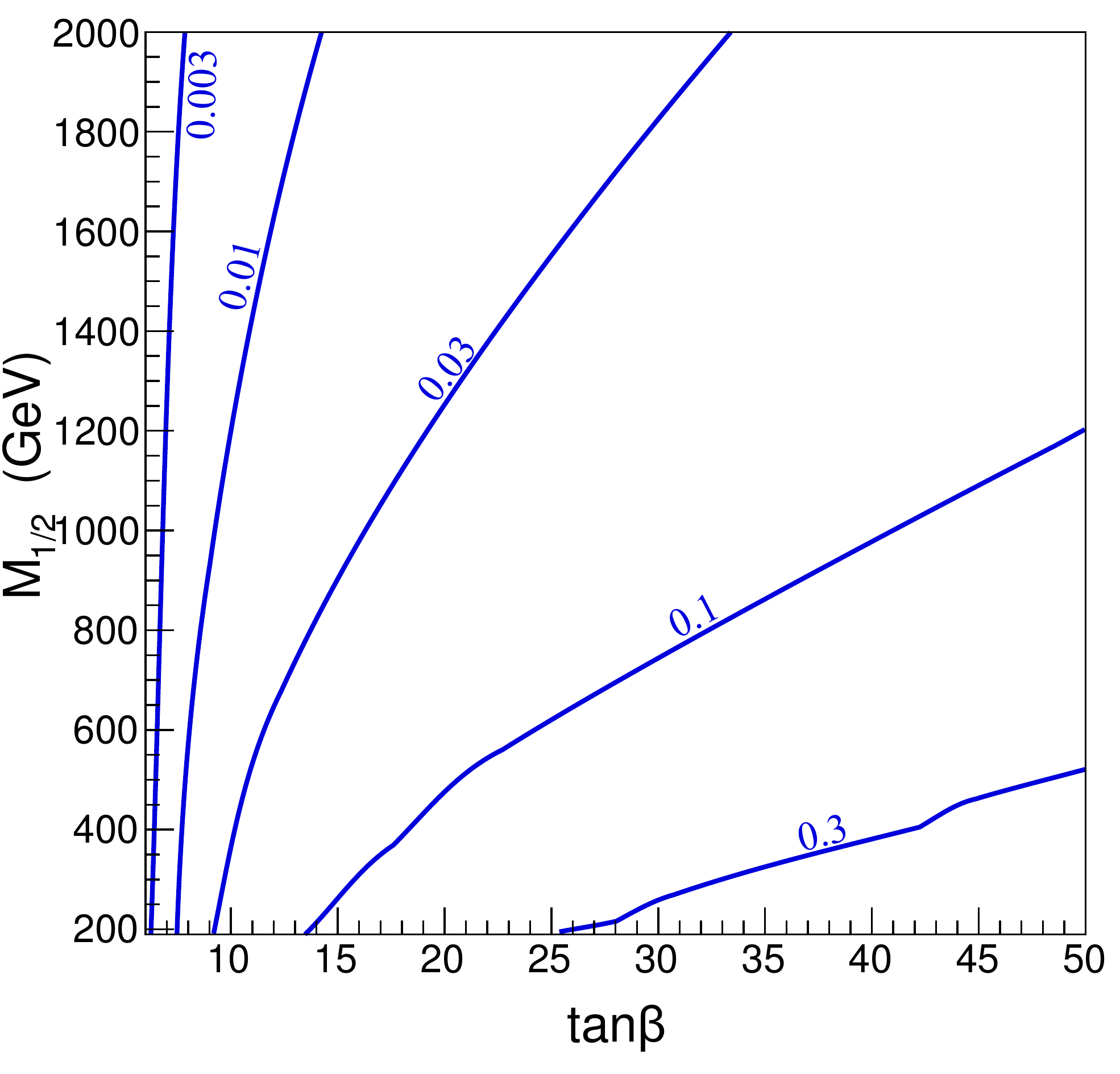}
    \label{fig:gmu}
\caption{Contours of the supersymmetric contribution to $(g-2)_\mu$ in
  units of $10^{-9}$.  Every point in the parameter space is in the FP
  region and satisfies $\Omegachi \simeq 0.23$, $A_0 = 0$, and $\mu >
  0$.}
\label{fig:gmuon}
\end{figure}

As noted in \secref{FPSUSY}, however, FP SUSY is far more general than
the FP region.  In particular, in FP SUSY, the smuon and muon
sneutrino need not have masses unified with the other scalars, and so
may be much lighter than the third generation squarks.  To explore
this possibility and its implications for $(g-2)_{\mu}$, we consider
the slight modification of mSUGRA/CMSSM in which all scalars have
GUT-scale mass $m_0$, except for the smuons and muon sneutrino.  This
modification is intended to be schematic, demonstrating the behavior
of $(g-2)_\mu$ with lowered smuon masses without bias toward a
particular approach.  A fully consistent approach must consider flavor
and GUT unification issues.  For simplicity, we take the smuon masses
to be degenerate at the weak-scale, with physical masses
\begin{equation}
M_{\tilde{\mu}} \equiv m_{\tilde{\mu}_L} = m_{\tilde{\mu}_R} =
m_{\tilde{\nu}_\mu}\ .
\end{equation}
At each point in the $(\mgaugino, \tan\beta)$ plane, we determine the
value of $M_{\tilde{\mu}}$ that gives $\Delta (g-2)_\mu^{\text{SUSY}}$
that either brings the theoretical prediction into complete agreement
with the central experimental value or reduces the discrepancy to
$2\sigma$.  Note that the dominant factor in the determination of the
relic density is the Higgs soft mass, with the sfermion masses
providing subleading effects, as long as $m_{\tilde{q}} \agt 500~\gev$
and $m_{\tilde{\ell}} \agt 200~\gev$~\cite{Feng:2010ef}.  The smuons
can therefore be quite light without affecting the relic density
constraint.

The results are given in \figref{msmu}.  As $\mgaugino$ increases, the
required smuon mass decreases to maintain a constant SUSY contribution
to $(g-2)_{\mu}$, and at some point, the required $M_{\tilde{\mu}}$
becomes too low, as it implies a $\tilde{\mu}$
LSP.\footnote{$M_{\tilde{\mu}}$ is cut off at $1.1 \times m_\chi$
numerically to avoid recalculating the relic density due to
$\tilde{\mu}-\chi^0$ coannihilation.}  The supersymmetric contribution
$\Delta (g-2)_\mu^{\text{SUSY}}$ also has a linear dependence on
$\tan\beta$, and so at large $\tan\beta$, there are allowed solutions
for larger values of $\mgaugino$ and $M_{\tilde{\mu}}$.

\begin{figure}
  \subfigure[\ $M_{\tilde{\mu}}$ (GeV) required to eliminate the
    discrepancy in $(g-2)_\mu$]{
    \includegraphics[width=.48\textwidth]{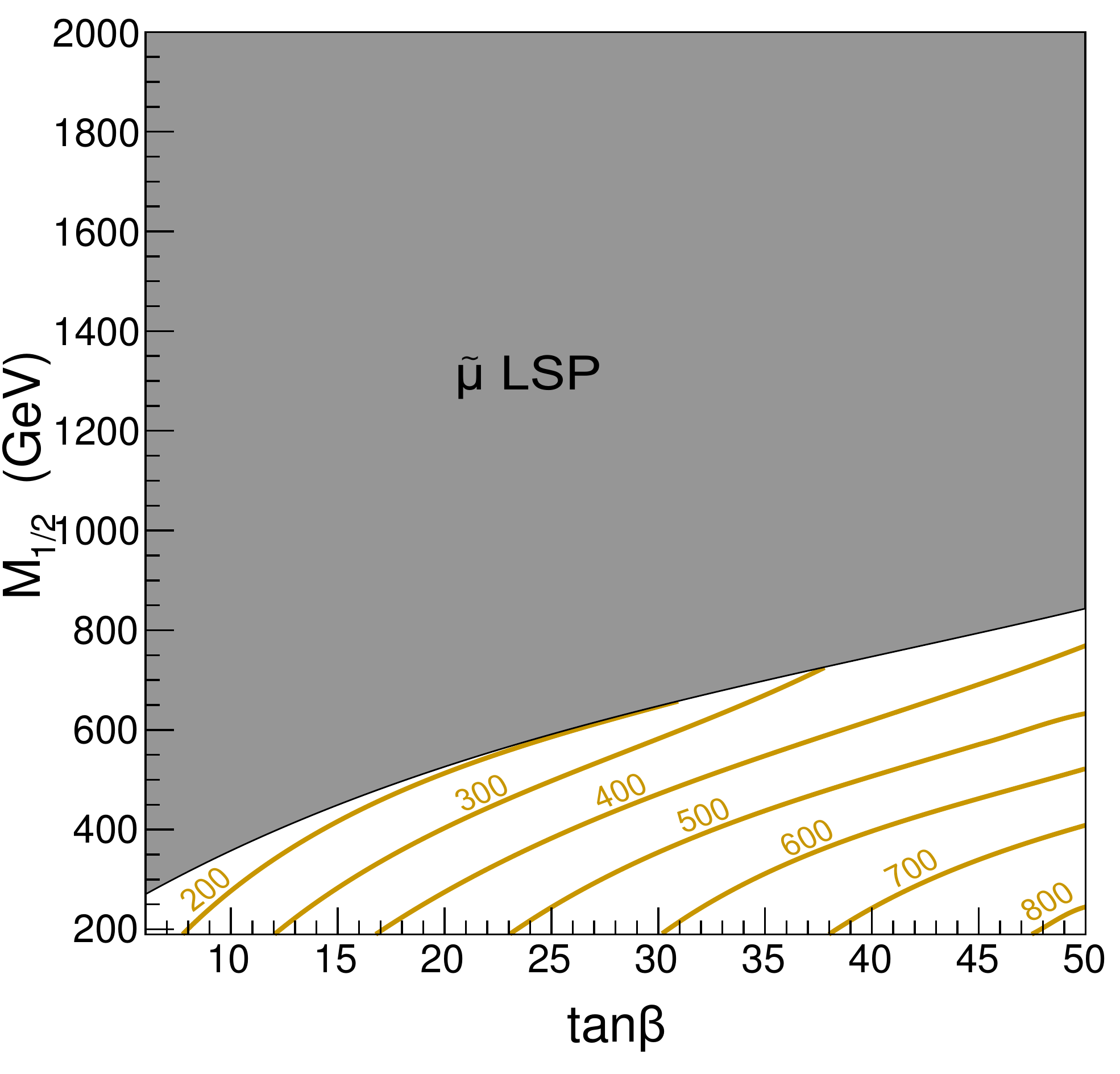}
    \label{fig:msmu_central} }
  \subfigure[\ $M_{\tilde{\mu}}$ (GeV) required to reduce the
    discrepancy in $(g-2)_\mu$ to $2\sigma$]{
    \includegraphics[width=.48\textwidth]{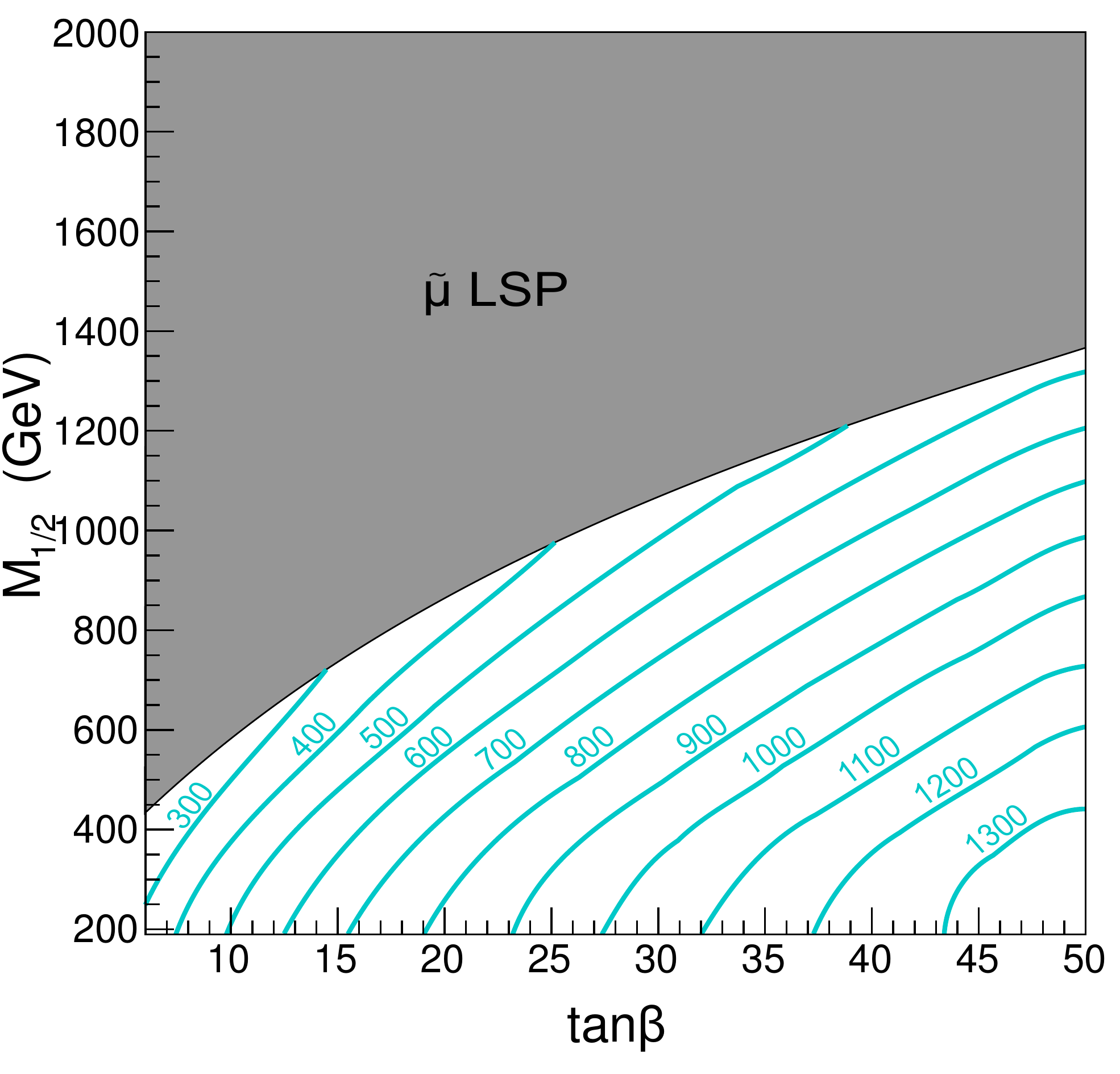}
    \label{fig:msmu_2sigma} }
\caption{Contours of $M_{\tilde{\mu}}$ required to eliminate the
  discrepancy between the theoretical and experimental values for
  $(g-2)_\mu$ (left) and to reduce the discrepancy to $2\sigma$
  (right). This model framework is a slight modification of
  mSUGRA/CMSSM in which all scalars have GUT-scale mass $m_0$, except
  for the smuons and muon sneutrino, which have physical mass
  $M_{\tilde{\mu}}$. In the shaded regions, the $\tilde{\mu}$ becomes
  the LSP. To specify all parameters aside from the smuon and muon
  sneutrino masses, every point in the parameter space is in the FP
  region and satisfies $\Omegachi \simeq 0.23$, $A_0 = 0$, and $\mu >
  0$.}
\label{fig:msmu}
\end{figure}

It is important to check that the scenarios for resolving the
$(g-2)_\mu$ discrepancy are viable in light of null results from LHC
new physics searches.  The model-independent bounds on slepton masses
are, of course, far weaker than those on squark masses.  The best
limits on slepton masses are still those from LEP2, which require
$m_{\tilde{\mu}} \agt 100~\gev$~\cite{Feng:2009te}.  In the future,
with $30~\ifb$ of data at 14 TeV, the LHC will be able to discover
sleptons through Drell-Yan production for $m_{\tilde{\mu}_L} \alt
300~\gev$ and $m_{\tilde{\mu}_R} \alt 200~\gev$~\cite{Andreev:2004qq}.
Greater sensitivity may be available in scenarios where the sleptons
are produced in cascades~\cite{Eckel:2011pw}.  However, in the FP
region where all other scalars are heavy and gluino production
dominates, if the sleptons are heavier than all charginos and
neutralinos, they will not be produced in gluino cascades, and so the
Drell-Yan limits apply.  This is the case for regions of the
$(\tan\beta, \mgaugino)$ plane shown in \figref{msmu}, and so there
are viable FP SUSY scenarios that resolve the $(g-2)_\mu$ discrepancy.
It would, however, be interesting to investigate scenarios motivated
by the $(g-2)_\mu$ discrepancy in which sleptons are produced in
gluino cascades.

\section{Conclusions}
\label{sec:conclusions}

SUSY models with heavy squarks and sleptons have long been motivated
by constraints on flavor- and CP-violation, the LEP2 constraint on the
Higgs boson mass, and other constraints, such as proton decay bounds.
Recent null results from LHC SUSY searches have further focused
attention on this possibility, and the interest in such scenarios is
especially heightened by the currently allowed Higgs boson mass window
$115.5~\gev < m_h < 127~\gev$, and tentative indications from the
ATLAS and CMS experiments for a Higgs boson with mass near 125 GeV.

Generic SUSY scenarios with heavy sfermions, and particularly heavy
top and bottom squarks, imply fine-tuning of the weak scale,
subverting the basic motivation for weak-scale SUSY.  In FP SUSY,
however, this is not the case.  The mass parameter $m_{H_u}^2$ evolves
to values around $m_Z^2$ at the weak scale, almost independent of its
GUT-scale starting value.  This focusing of RG trajectories implies
that the weak scale in FP SUSY theories is not fine-tuned with respect
to variations in the fundamental SUSY-breaking parameters.  Note that
the fact that $m_{H_u}^2$ evolves to values around $m_Z^2$ at the weak
scale for a particular choice of GUT-scale parameters is necessary to
remove fine-tuning with respect to variations in $\mu$, and is
possible for other choices of GUT-scale parameters (see, for example,
Ref.~\cite{Horton:2009ed,Feldman:2011ud}).  However, naturalness with
respect to variations in {\em all} SUSY-breaking parameters requires
that $m_{H_u}^2$ evolve to a weak-scale value irrespective of its
starting value, and so the focus point behavior of renormalization
group trajectories is an essential feature of any natural theory with
multi-TeV top and bottom squarks motivated by the currently allowed
Higgs boson mass range.

In this study, we have focused for the most part on models of FP SUSY
that are also part of the mSUGRA/CMSSM framework.  These FP region
models naturally produce Higgs boson masses above the LEP2 bound of
114.4 GeV, and suppress electron and neutron EDMs sufficiently, even
for ${\cal O}(1)$ phases.  To more globally display the predictions of
FP SUSY, we have required $\Omegachi \simeq 0.23$ and plotted results
in the $(\tan\beta, \mgaugino)$ plane.  We find that FP SUSY naturally
accommodates Higgs boson masses up to 120-124 GeV, which, given an
estimated 2 GeV uncertainty in the theoretical calculation, is
consistent with current Higgs boson mass indications.  In addition, we
have shown that FP SUSY is naturally consistent with constraints from
$b \to s \gamma$, $B_s \to \mu^+ \mu^-$, and null results from dark
matter direct detection experiments.  Finally, in general FP SUSY with
a non-unified smuon mass, we have found that FP SUSY may resolve the
discrepancy in $(g-2)_{\mu}$ consistent with all current constraints.

Given these successes, it is natural to ask what evidence for FP SUSY
should accumulate in the near future if FP SUSY is realized in nature.
Certainly the Higgs boson should be discovered with a mass in the
currently allowed mass window, and searches for SUSY from gluino pair
production, followed by gluinos cascading through charginos and
neutralinos are promising for some of the parameter
space~\cite{Chattopadhyay:2000qa,Mercadante:2005vx,DeSanctis:2007td,%
Das:2007jn,Kadala:2008uy}. Equally exciting would be the discovery of
dark matter with a spin-independent $\chi$-nucleon cross section near
the zeptobarn scale, which is a robust prediction of mixed
Higgsino-Bino dark matter with heavy squarks and sleptons. Finally,
most signals of indirect dark matter detection are also generically
enhanced in the FP SUSY scenario~\cite{Feng:2000zu}.

\section*{Acknowledgments}

We thank Daniel Feldman and AseshKrishna Datta for discussions and
Wonsang Cho for collaboration in the early stages of this work.  The
work of J.L.F.~and D.S.~was supported in part by NSF grant
PHY--0970173.  The work of K.T.M.~was supported in part by DOE grant
DE--FG02--97ER41029.  The work of D.S.~was supported in part by a UC
Irvine Graduate Dean's Dissertation Fellowship.

\appendix

\section*{Appendix}
\label{sec:appendix}

It is well known that different spectrum calculators do not give
identical results for the SUSY mass spectrum, even for the same set of
input parameters~\cite{Allanach:2003jw,Belanger:2005jk}. The reasons
for this apparent discrepancy are well understood; see
Ref.~\cite{Baer:2005ky} for a nice summary. Above all, one should keep
in mind that the SUSY spectrum is always calculated at a fixed order
in perturbation theory, and there is an intrinsic uncertainty due to
neglecting the higher order terms in perturbation theory.  The main
differences between the various programs arise mostly because they
choose to neglect different sets of higher-order terms.  For example,
one may choose to use either tree-level or 1-loop-corrected masses in
the radiative corrections, or choose a slightly different value for
the matching scale between the SM and the MSSM. In each case, the
difference between the two options is a higher-order effect. In this
paper, we chose to work with the SOFTSUSY program, but we expect
qualitatively similar results from other spectrum generators as well.

\begin{figure}[t]
    \includegraphics[width=.7\textwidth]{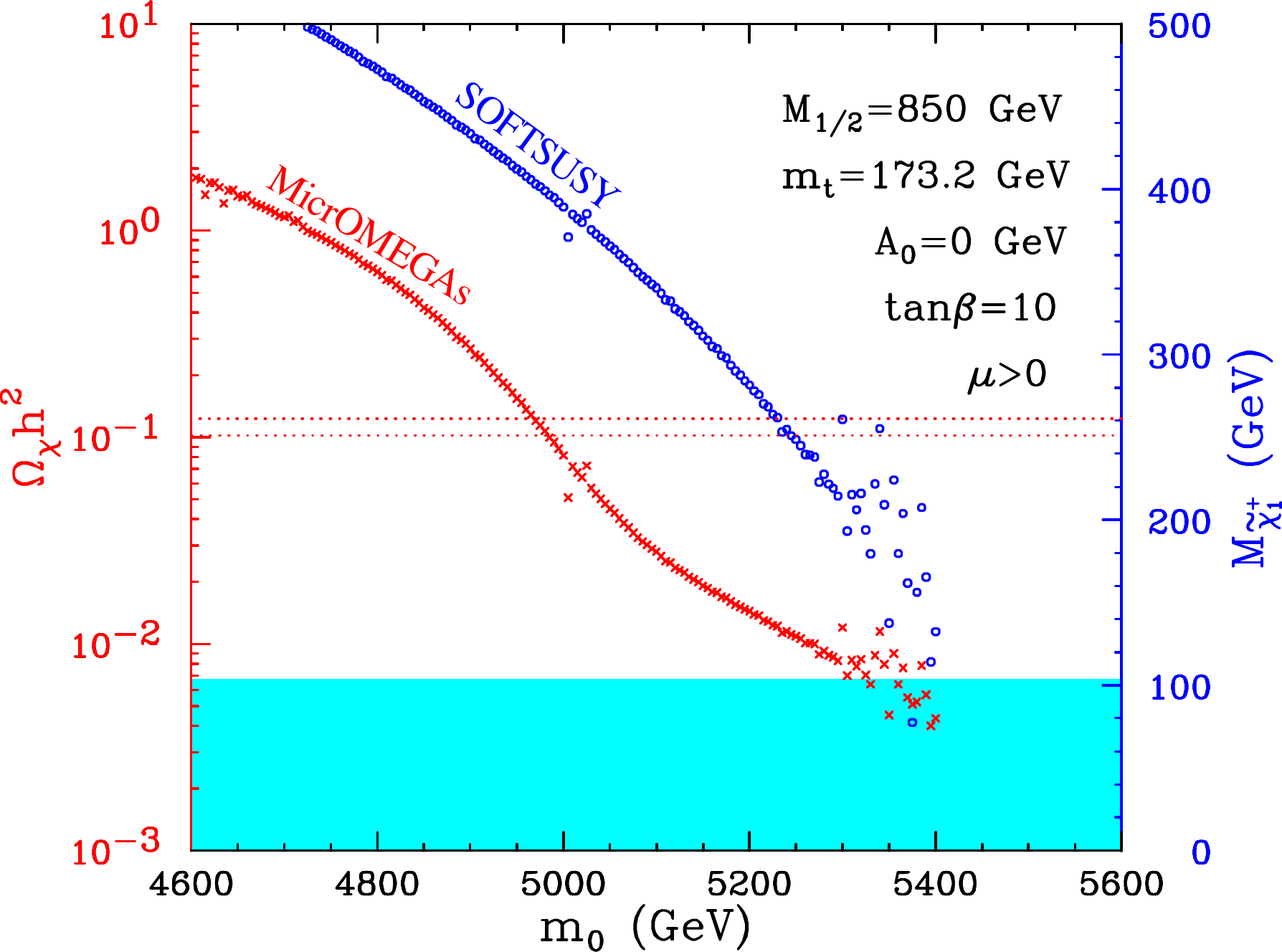}
\caption{A slice through the mSUGRA parameter space from
  \figref{EDMtanb10} for a fixed $\mgaugino = 850~\gev$, showing
  results for the chargino mass $M_{\tilde\chi^+_1}$ from SOFTSUSY
  (blue dots) and for $\Omega_\chi h^2$ from MicrOMEGAs (red crosses).
  The cyan shaded region is excluded by chargino searches at LEP, and
  the horizontal dotted lines mark the $3\sigma$ preferred region for
  $\Omegachi h^2$. }
\label{fig:omega_m0}
\end{figure}

On a related topic, each spectrum calculator needs to solve a
two-sided boundary value problem, since the boundary conditions for
the gauge and Yukawa couplings are specified at the weak scale, while
the soft SUSY-breaking parameters are given at the (yet to be
determined) GUT scale. The standard approach used by all programs is
to apply iterations until converging on a solution.  Unfortunately, on
occasion one may encounter poor convergence as a sign of a chaotic
behavior~\cite{Matchev:2012vf}.  This is illustrated in
\figref{omega_m0}, which takes a slice through the $(m_0, \mgaugino)$
plane of \figref{EDMtanb10} in 5 GeV increments along $m_0$, for a
fixed value of $\mgaugino = 850~\gev$.  The figure shows the chargino
mass $M_{\tilde\chi^+_1}$ calculated by SOFTSUSY (right axis) and the
relic abundance calculated by MicrOMEGAs (left axis).  We see that at
low $m_0$, SOFTSUSY is able to converge, and both quantities follow a
well-defined trend. However, at sufficiently large values of $m_0$,
SOFTSUSY is not able to achieve the desired level of convergence, and
the obtained results (upon exiting after a fixed number of iterations)
visibly deviate from the expected trend. As seen in \figref{omega_m0},
in principle this presents a problem for the correct mapping of the
boundary of the region allowed by LEP chargino searches
($M_{\tilde\chi^+_1} > 103~\gev$). Fortunately, however, the parameter
space points with the desired value of the relic density ($\Omegachi
h^2\approx 0.1$) are relatively safe, since they are still well within
the region with good convergence, and the maps shown in
\figsref{m0mu}{mhiggsmchitanbeta} are robust.

\bibliography{bibfpreduxprd}{}

\end{document}
